\documentclass[nofootinbib,showkeys]{revtex4}%
\usepackage{etoolbox}
\usepackage[colorlinks=true]{hyperref}
\AtBeginDocument{
	\hypersetup{
		urlcolor=cyan,
		citecolor=magenta,
		linkcolor=blue,
		anchorcolor=green,
	}%
}
\usepackage[colorinlistoftodos]{todonotes}

\usepackage{amssymb,yfonts}
\usepackage{amsmath}
\usepackage{extpfeil}
\usepackage{mathtools}
\usepackage{amsfonts}
\usepackage{graphicx}
\usepackage{ulem} 
\usepackage{cancel}
\usepackage{epstopdf}
\usepackage{extarrows}
\usepackage{multirow}
\usepackage{enumitem}

\usepackage{tikz}
\usetikzlibrary{arrows.meta, calc, positioning}
\usetikzlibrary{matrix,arrows}

\providecommand{\mtn}[1]{{\scriptscriptstyle \!\, #1}}%
\definecolor{lred}{rgb}{1,0.3,0}
\definecolor{dgreen}{rgb}{0.0,0.6,0}

\def\ds{\displaystyle}

\def\md{\mathrm{d}}

\definecolor{amaranth}{rgb}{0.9, 0.17, 0.31}
\definecolor{purple(munsell)}{rgb}{0.62, 0.0, 0.77}
\definecolor{americanrose}{rgb}{1.0, 0.01, 0.24}
\definecolor{palatinateblue}{rgb}{0.15, 0.23, 0.89}
\definecolor{royalblue(web)}{rgb}{0.25, 0.41, 0.88}
\definecolor{hanpurple}{rgb}{0.32, 0.09, 0.98}
\definecolor{beaublue}{rgb}{0.74, 0.83, 0.9}
\definecolor{carminered}{rgb}{1.0, 0.0, 0.22}
\definecolor{brightpink}{rgb}{1.0, 0.0, 0.5}
\definecolor{vividviolet}{rgb}{0.62, 0.0, 1.0}
\hypersetup{ linktoc=all,
	colorlinks, linkcolor={palatinateblue},
	citecolor={brightpink}, urlcolor={amaranth}}

\def\sideremark#1{\ifvmode\leavevmode\fi\vadjust{\vbox to0pt{\vss
			\hbox to 0pt{\hskip\hsize\hskip1em
				\vbox{\hsize2cm\tiny\raggedright\pretolerance10000
					\noindent #1\hfill}\hss}\vbox to8pt{\vfil}\vss}}}%

\makeatletter
\newcommand{\qbar}{\text{\q@bar}}
\newcommand{\q@bar}{%
	\vphantom{$\m@th q$}%
	\ooalign{%
		$\m@th q$\cr
		\hidewidth\kern.15em\smash{\raisebox{-0.7ex}{$\m@th\mathchar'26$}}\hidewidth\cr}%
}
\makeatother

\begin{document}
\title[]{Newtonian and Relativistic Particle Mechanics from Local Energy Conservation: A Symmetry-Selected Force–Energy Framework}
\author{Thomas Oikonomou}
\email{thomas.oikonomou1@gmail.com}
\affiliation{Institute of Nanoscience and Nanotechnology, National Center for Scientific Research “Demokritos”, 15310 Athens, Greece}
%
%
%

%


\begin{abstract}

We develop a formulation of particle mechanics in which the functional relation between force and kinetic energy is derived directly from the instantaneous (pointwise along the trajectory) conservation of mechanical energy $E$, rather than postulated through Newton’s second law or a variational principle. Starting from the condition $\dot{E}=0$, we derive a generalized force law that does not assume a specific kinetic-energy function, momentum–velocity relation, or equation of motion. The resulting inertial response naturally decomposes into a component parallel to the acceleration, responsible for changes in kinetic energy, and a transverse component that preserves energy while altering the direction of motion.
Imposing rotational equivariance constrains the geometric structure of the force law, while the relativity principle between inertial reference frames further restricts its admissible realizations. In strictly one-dimensional motion, inertial frame equivalence, supplemented by a set of five structural properties, implies invariance of the total force under inertial boosts; together with local energy conservation, this uniquely fixes the functional form of kinetic energy and momentum. Galilean invariance selects the Newtonian expressions, whereas Lorentz invariance yields the relativistic ones.
The framework unifies conservative and non-conservative (yet non-dissipative, e.g., the magnetic Lorentz force) dynamics at the single-particle level and clarifies the precise conditions under which energy-based and variational formulations of mechanics are dynamically equivalent. Newtonian and relativistic mechanics thus emerge as symmetry-selected realizations of a common energy-conserving force-energy structure. The analysis further shows that the standard Lagrangian form $L=T-V$ is consistent with, and implicitly presupposes, Galilean relativity.

\end{abstract}

\eid{ }
\date{\today }
\startpage{1}
\endpage{1}
\maketitle

\section{Introduction}

Classical mechanics of a massive particle, described by its mass $m$, position $\mathbf{x}$, velocity $\mathbf{v}=\dot{\mathbf{x}}$, and acceleration $\mathbf{a}=\dot{\mathbf{v}}$ in three-dimensional Euclidean space, rests on two closely related but conceptually distinct foundations: Newton’s laws of motion and the Principle of Least Action (PLA). In the variational formulation, the equations of motion follow from extremizing an action functional, with conservation laws emerging via Noether’s theorem \cite{Feynman1948,Lanczos1986,Arnold1989,Brizard2008,Feynman2010,Rojo2018,Franklin2017,Wen2023,Neother1918}. In the force-based formulation, dynamics is postulated directly through Newton’s second force law, $\mathbf{F}=m\mathbf{a}$, with energy conservation appearing as a derived consequence.

A nontrivial structural feature underlies both approaches. In Newtonian mechanics, the quadratic kinetic energy $T=\frac{1}{2}mv^2$ follows from the work–energy theorem after postulating the linear force–acceleration relation $\mathbf{F}=m\mathbf{a}$. Conversely, within the PLA, the same quadratic form must be specified a priori in the Lagrangian in order to recover Newton’s second law. Thus, the Newtonian force law and the Newtonian kinetic energy do not arise independently, but as a mutually consistent pair $\{T,\mathbf{F}\}$, selected by shared structural assumptions rather than deduced from a deeper principle.

This reciprocity raises a foundational question: is the familiar pairing between force and kinetic energy a fundamental necessity, or merely one realization among a broader class of dynamically consistent force–energy relations? If the latter is the case, then classical mechanics should admit a formulation in which force laws and kinetic-energy functions emerge jointly from primitive principles, rather than being separately postulated.
Aim of this paper is to address this question. To this end, we adopt mechanical energy conservation as the primary organizing principle from which the dynamical structure of the theory is to be inferred.

The idea of elevating energy conservation to a foundational role in mechanics has been explored from several perspectives \cite{Lindgren2002,Hanc2004,Vinokurov2014,Carlson2016,Baumgarten2024A,Zhou2022,Baumgarten2024,Goyal2020}. For example, Zhou and Wang \cite{Zhou2022} proposed a reformulation of analytical mechanics in which energy conservation is taken as the primary axiom. Within this framework, Lagrange’s equations, Hamilton’s equations, and Hamilton–Jacobi theory are systematically recovered, and standard force-based and variational formulations emerge as consequences of an underlying energetic balance. However, the Newtonian kinetic-energy form $T=\frac{1}{2}mv^2$ is assumed a priori. As a result, the question of whether the functional form of kinetic energy can be fixed independently of a specific inertial-response model remains open.

A closely related perspective is developed by Baumgarten \cite{Baumgarten2024}, who explicitly adopts mechanical energy conservation as the foundational principle and assumes an additive decomposition $E(x,v)=T(v)+V(x)$ in one-dimensional motion. Newtonian dynamics and the quadratic kinetic-energy form are then recovered directly from the local condition $\dot{E}=0$, making their mutual interdependence explicit. Extensions to relativistic dynamics are discussed in a momentum-based formulation, but the relativistic energy–momentum relation is taken as input rather than derived, so the structure of the relativistic force–energy pair remains unaddressed.

Earlier works adopting energy-first perspectives \cite{Hanc2004,Carlson2016,Baumgarten2024A,Vinokurov2014} pursue related conceptual directions, but primarily aim to reconstruct known dynamical frameworks rather than determine whether the force–energy relation itself can be fixed directly from local energy conservation without presupposing a specific inertial model.

In a related spirit, Goyal \cite{Goyal2020} investigates whether the functional form of the pair 
$\{T,\mathbf{F}\}$ can be fixed by conservation principles and Inertial Reference Frame (IFR) symmetries alone. By analyzing idealized elastic collisions and imposing frame invariance, the author  derives the nonrelativistic and relativistic kinetic-energy functions and the corresponding momentum. The dynamical force law is introduced subsequently via a staccato model of motion, in which continuous dynamics are represented as sequences of infinitesimal velocity changes whose rates are additionally assumed to depend on the masses, positions, and velocities of the interacting bodies. While this approach is general in scope, it does not address the specific question posed in the present work, as the coordinate transformation, Galilean or Lorentz, is assumed from the outset and the corresponding dynamical equations are then derived within each case.

The present work develops a purely local, single-particle route to mechanics at the level of forces and kinetic energy. Starting from the instantaneous conservation of mechanical energy, $\dot{E}=0$, imposed as a pointwise constraint along the particle trajectory, we derive a generalized force law that links kinetic energy and force without assuming a specific kinetic-energy form, momentum–velocity relation, or equation of motion. Rotational symmetry further constrains the admissible structure of the force law, while the Relativity Principle (RP) between IRFs selects its explicit realizations: Galilean invariance yields the Newtonian expressions, and Lorentz invariance yields the relativistic ones. In this way, the relation between kinetic energy and force is established prior to any kinematical assumptions, and the role of the RP is confined to selecting among dynamically consistent realizations rather than fixing the structure of the force law itself.

Compared to Ref. \cite{Baumgarten2024}, the novelty of the present work lies in the fact that applying the condition $\dot{E}=0$ in three spatial dimensions yields a generalized force law that naturally includes a transverse component involving a cross product. This additional structure resolves the limitation identified in Ref. \cite{Baumgarten2024}, which arises from its restriction to one spatial dimension. At the methodological level, a complementary distinction arises relative to Goyal’s energetic reconstruction programme. Unlike his collision-based approach, the present work derives the local force–energy structure directly from pointwise energy conservation along a single trajectory, establishes the generalized inertial response prior to any kinematical specialization, and determines the precise structural conditions under which this framework coincides with or extends beyond the standard variational formulation.

The structure of the paper reflects the logical hierarchy of principles underlying the proposed framework. 
In Sec. \ref{SecII}, we take the instantaneous form of the Principle of Conservation of Energy (PCE), $\dot{E}=0$, as a pointwise constraint along the motion of a single particle and derive from it a generalized force law. This force law is obtained without assuming any specific kinetic-energy function, momentum–velocity relation, or equation of motion. Its structure naturally decomposes the inertial response into a component collinear with the acceleration, responsible for kinetic-energy change, and a transverse, energy-preserving component that modifies only the direction of motion. By further imposing $SO(3)$-equivariance \cite{Villar2023}, we determine the most general vectorial form of the associated momentum and inertial-response vectors compatible with spatial isotropy, and show that all admissible expressions are governed by a single scalar function of the speed, $f(v):=v^{-1}\partial T(v)/\partial v$.

In Sec.~\ref{SecIII}, we reduce the particle dynamics to strictly one-dimensional (1D) motion and show that the Relativity Principle (RP), together with a minimal set of physically natural structural requirements on force transformations, implies invariance of the total force under inertial boosts. This result, referred to as one-dimensional Force-Boost Invariance (1D-FBI), is a purely kinematical consequence of IRF equivalence and does not presuppose any particular dynamical law.

In Sec.~\ref{SecIIIb}, we combine 1D-FBI with the kinematical relations between velocities and accelerations in different inertial frames, thereby fixing the explicit form of the function $f(v)$. Under Galilean transformations~\cite{Fitzpatrick2021}, $f(v)$ is constant, yielding the standard Newtonian expressions for kinetic energy, momentum, and force vectors. Under Lorentz transformations~\cite{Lorentz1904}, $f(v)$ is uniquely determined to produce the relativistic kinetic energy and Planck’s momentum, while simultaneously introducing an acceleration-dependent transverse inertial response. Notably, these results are obtained entirely within 3D Euclidean space, with time treated as an external evolution parameter, and without invoking variational principles, or a priori momentum–velocity relations~\cite{Catoni2011,Naber2012}. We remark however that although spacetime geometry is not invoked explicitly, the Lorentz kinematics underlying the formulation encode an equivalent geometric structure in implicit form.

In Sec.~\ref{SecIV}, we examine the relationship between the present energy-based formulation and PLA. By identifying the conserved Noether quantity with the mechanically defined energy, we derive the structural conditions under which a variational formulation reproduces the same dynamics as the PCE-derived force law. This analysis clarifies when PLA and PCE are dynamically equivalent and delineates the broader class of energy-conserving dynamics that lie beyond the standard variational framework. Conclusions are presented in Section~\ref{Conl}.

\section{Conservation of Energy and the respective Force Law}\label{SecII}
%
Consider a massive particle in a 3D Euclidean space, characterized by its position, $\mathbf{x}=\{x_1,x_2,x_3\}$, velocity, $\mathbf{v}=\{v_1,v_2,v_3\}$, and acceleration, $\mathbf{a}=\{a_1,a_2,a_3\}$, all expressed in Lagrangian coordinates, with  $\mathbf{a}(t) = \frac{\md}{\md t}\mathbf{v}(t) = \frac{\md^2}{\md t^2}\mathbf{x}(t)$, and time $t$ treated as an external evolution parameter.
We define its mechanical energy $E=E(\mathbf{x},\mathbf{v})$ \cite{Coelho2024,Cabaret2024} as the sum of a velocity-dependent kinetic energy $T(v)$ and a position-dependent potential energy $V(\mathbf{x})$,
\begin{eqnarray}\label{eqII4}
    E= T(v)+V(\mathbf{x})\,,
\end{eqnarray}
where $v=\|\mathbf{v}\|=\sqrt{v_1^2+v_2^2+v_3^2}$ is the magnitude of the velocity vector. 
We assume that $E$ is separable into position- and velocity-dependent parts, without explicit time dependence, and that $T$ depends only on the speed $v$.

The decomposition in Eq.~\eqref{eqII4} constitutes a structural assumption that restricts the class of systems under consideration, excluding systems with velocity-dependent or time-varying potentials (as in electromagnetism), and dissipative systems in which mechanical energy is not conserved. The present work is intentionally confined to this separable setting, as it provides the minimal framework in which a well-defined notion of mechanical energy exists and its conservation can be imposed locally through the condition $\dot{E}=0$. This class includes a wide range of standard mechanical systems, such as particles in conservative force fields (e.g. the harmonic oscillator), for which the decomposition in Eq.~\eqref{eqII4} provides a consistent and physically meaningful description.

Taking the time derivative of the energy gives
\begin{eqnarray}
\dot{E}
&=&
\frac{\md T(v)}{\md t} + \frac{\md V(\mathbf{x})}{\md t}
=
\sum_{i=1}^{3} \frac{\partial T(v)}{\partial v_i}\frac{\md v_i}{\md t}
+
\sum_{i=1}^{3} \frac{\partial V(\mathbf{x})}{\partial x_i}v_i
=
\sum_{i=1}^{3} 
v_i\left[
\left(\frac{1}{v_i}\frac{\partial T(v)}{\partial v_i}\right)\frac{\md v_i}{\md t}
+
\frac{\partial V(\mathbf{x})}{\partial x_i}
\right]\,.
\end{eqnarray}
The kinetic term $\frac{1}{v_i}\frac{\partial T(v)}{\partial v_i}$ simplifies using the chain rule
\begin{eqnarray}\label{fdef}
\frac{1}{v_i}\frac{\partial T(v)}{\partial v_i}
=
2\frac{\partial T(v)}{\partial v_i^2}
=
2\frac{\partial T(v)}{\partial v^2}\frac{\partial v^2}{\partial v_i^2}
=
2\frac{\partial T(v)}{\partial v^2}
=
\frac{1}{v}\frac{\partial T(v)}{\partial v}
=:f(v)\,.
\end{eqnarray}
Substituting this into Eq.~\eqref{fdef}, the time derivative of the energy can be expressed as the following dot product
\begin{eqnarray}\label{EnergyRoC1}
\dot{E}
=
\sum_{i=1}^{3} 
v_i\left[
f(v)\frac{\md v_i}{\md t}
+
\frac{\partial V(\mathbf{x})}{\partial x_i}
\right]
=
\mathbf{v}\cdot \mathbf{n}\,,\qquad
\mathbf{n}:=
f(v)\mathbf{a}
+
\boldsymbol{\nabla}_\mathbf{x} V(\mathbf{x})\,,
\end{eqnarray}
where $\boldsymbol{\nabla}_\mathbf{x}:=(\frac{\partial}{\partial x_1},\frac{\partial}{\partial x_2},\frac{\partial}{\partial x_3})^T$, and $\mathbf{v}\cdot\mathbf{a}=\sum_i v_i \frac{\md v_i}{\md t}$.

Exploring the conditions for energy conservation ($\dot{E}=0$), we identify three distinct possibilities from Eq. (\ref{EnergyRoC1}),
i. $\mathbf{v} = \mathbf{0}$,
ii. $\mathbf{v} \cdot \mathbf{n} = 0$, and
iii. $\mathbf{n} = \mathbf{0}$.
The first case, $\mathbf{v} = \mathbf{0}$, corresponds to trivial static motion and is not of dynamical interest. The second condition, $\mathbf{v} \cdot \mathbf{n} = 0$, implies that $\mathbf{v} \perp \mathbf{n}$. In three spatial dimensions, any vector orthogonal to $\mathbf{v}$ can be expressed as a cross product of the form $\mathbf{n} = \mathbf{v} \times \boldsymbol{\Theta}$, for some vector $\boldsymbol{\Theta} = \boldsymbol{\Theta}(\mathbf{x}, \mathbf{v}, \mathbf{a})$. This follows from the fact that the mapping $\boldsymbol{\Theta}\mapsto\mathbf{v}\times \boldsymbol{\Theta} $ generates vectors orthogonal to $\mathbf{v}$, and in three dimensions its image coincides with the full subspace perpendicular to $\mathbf{v}$. The third condition, $\mathbf{n} = \mathbf{0}$, is a special case of the second, occurring when $\boldsymbol{\Theta} \parallel \mathbf{v}$ or $\boldsymbol{\Theta} = \mathbf{0}$. Thus, the most general requirement for $\dot{E}=0$ is condition (ii), i.e. $\mathbf{n} = \mathbf{v} \times \boldsymbol{\Theta}$.

At this stage, the condition $\mathbf{n}=\mathbf{v}\times\boldsymbol{\Theta}(\mathbf{x},\mathbf{v},\mathbf{a})$  specifies a general structure compatible with local energy conservation, but does not uniquely fix the functional dependence of the vector  $\boldsymbol{\Theta}$. We therefore assume $\boldsymbol{\Theta}=\mathbf{K}(\mathbf{x},\mathbf{v})-\mathbf{M}(\mathbf{x},\mathbf{v},\mathbf{a})$, whose purpose is to separate contributions that are independent of acceleration from those that depend explicitly on it. The term 
$\mathbf{K}(\mathbf{x},\mathbf{v})$ collects all components that can be interpreted as externally prescribed influences (including velocity-dependent forces), while $\mathbf{M}(\mathbf{x},\mathbf{v},\mathbf{a})$ encodes the part of the response intrinsically tied to acceleration. We emphasize that this decomposition is not unique. In particular, the transformation 
$\mathbf{K}\to \mathbf{K}+\boldsymbol{\Delta}(\mathbf{x},\mathbf{v})$, $\mathbf{M}\to\mathbf{M}+\boldsymbol{\Delta}(\mathbf{x},\mathbf{v})$, with arbitrary $\boldsymbol{\Delta}(\mathbf{x},\mathbf{v})$, leaves $\boldsymbol{\Theta}$ unchanged. The present choice should therefore be understood as an ansatz that isolates all acceleration dependence into $\mathbf{M}$, rather than as a unique or general decomposition. Under this assumption, the condition for energy conservation becomes
\begin{eqnarray}\label{GNLaw}
\mathbf{F}_\mathrm{c}(\mathbf{x}) + \mathbf{F}_{\perp}(\mathbf{x},\mathbf{v}) =f(v)\,\mathbf{a}+\mathbf{v}\times\mathbf{M}(\mathbf{x},\mathbf{v},\mathbf{a})\,,
\end{eqnarray}
where $\mathbf{F}_\mathrm{c}(\mathbf{x}) := -\boldsymbol{\nabla}_\mathbf{x} V(\mathbf{x})$ is a conservative force, and $\mathbf{F}_{\perp}(\mathbf{x}, \mathbf{v}) := \mathbf{v} \times \mathbf{K}(\mathbf{x}, \mathbf{v})$ is a transverse force orthogonal to the instantaneous velocity.

The l.h.s. of Eq.~\eqref{GNLaw} thus represents the applied forces, while the r.h.s. encodes acceleration–dependent response terms. Eq.~\eqref{GNLaw} therefore constitutes the most general local relation between the kinetic–energy function $T(v)$ and the total force $\mathbf{F}_\mathrm{tot}$ that follows directly from PCE, without invoking additional dynamical postulates. This equation is a central result of the present work. It establishes an explicit relation between the kinetic–energy function $T(v)$ and the total force through the scalar response function $f(v)$, thereby linking inertial response and energy change at a local level. Unlike Newton’s second law, which presupposes a linear relation between force and acceleration and therefore fixes the form of $T(v)$, the present framework imposes no prior restriction on 
$T(v)$. Velocity-dependent and acceleration-related contributions emerge naturally, yielding a richer dynamical structure in which Newtonian and relativistic mechanics appear as particular realizations determined by different admissible forms of $f(v)$.

The transverse force $\mathbf{F}_\perp=\mathbf{v}\times\mathbf{K}$ shares the geometric properties of the magnetic component of the Lorentz force: it is always perpendicular to the velocity, performs no work, and thus modifies only the trajectory without changing the kinetic energy. This shows that non-dissipative yet non-conservative forces arise naturally within the present formalism.

The term $\mathbf{v}\times\mathbf{M}$ is of a different nature. Its appearance follows necessarily from the structure of Eq.~\eqref{GNLaw}: being orthogonal to $\mathbf{v}$, it leaves the kinetic energy unchanged and therefore represents a purely inertial, energy–conserving modification of the motion. As demonstrated in Section~\ref{SecIII}, this term becomes essential in the relativistic regime, allowing Newtonian and relativistic dynamics to be embedded within a single Euclidean, energy–based framework.

By contrast, the response $f(v)\mathbf{a}$ plays a distinguished role: only its projection along the velocity contributes to the rate of change of kinetic energy, $\dot{T}=\mathbf{v}\cdot[f(v)\mathbf{a}]$, so that  acceleration alone mediates energy exchange.

To proceed, it is convenient to introduce the momentum vector $\mathbf{p}$ as a response vector whose time derivative equals the total applied force, $\mathbf{F}_\mathrm{tot} = \mathbf{F}_\mathrm{c} + \mathbf{F}_{\perp}=:\frac{\md}{\md t} \mathbf{p}$, so that Eq.~\eqref{GNLaw} may be written as
\begin{eqnarray}\label{GTMomentum}
\frac{\md\mathbf{p}}{\md t}=f(v) \mathbf{a} + \mathbf{v}\times\mathbf{M}\,.
\end{eqnarray}
At this stage, the momentum $\mathbf{p}$ is introduced as a definition, without assuming a priori any specific relation to the velocity. Its functional form will be determined later in the analysis.

So far, we have introduced all the quantities entering the force law. To further constrain the structure of the response functions $\mathbf{p}$ and $\mathbf{M}$, we now invoke rotational symmetry. Physical space is experimentally observed to be isotropic: rotating an experiment in the laboratory does not change its outcome. Accordingly, the momentum and force laws must retain their form under spatial rotations. Mathematically, this means that the vector quantities in the theory must transform covariantly, equivalently, be $SO(3)$-equivariant, under the rotation group. With this requirement, we can construct the most general $SO(3)$-equivariant forms of $\mathbf{p}$ and $\mathbf{M}$ from the available kinematic vectors ${\mathbf{x},\mathbf{v},\mathbf{a}}$. The details of the calculations below are presented in Appendix \ref{AppendixA0}.

If we consider  $\{\mathbf{x}, \mathbf{v}, \mathbf{a}\}$ as the set of the kinematic state vectors describing the 3D motion, and require that $\mathbf{p}=\mathbf{p}(\mathbf{x}, \mathbf{v}, \mathbf{a})$ and $\mathbf{M}=\mathbf{M}(\mathbf{x}, \mathbf{v}, \mathbf{a})$ are $SO(3)$-equivariant vector functions, then their most general, irreducible vector representation takes the  form \cite{Smith1971,Villar2023}
\begin{subequations}\label{MomentumDecomp}
\begin{eqnarray}
\label{MomentumDecomp_a}
\mathbf{p}
&=&
P_1\mathbf{x} + P_2\mathbf{v} + P_3\mathbf{a} + P_4(\mathbf{x}\times\mathbf{v}) + P_5(\mathbf{x}\times\mathbf{a}) + P_6(\mathbf{v}\times\mathbf{a})\,,\\
\label{MomentumDecomp_b}
\mathbf{M}
&=&
Q_1\mathbf{x} + Q_2\mathbf{v} + Q_3\mathbf{a} + Q_4(\mathbf{x}\times\mathbf{v}) + Q_5(\mathbf{x}\times\mathbf{a}) + Q_6(\mathbf{v}\times\mathbf{a})\,,
\end{eqnarray}
\end{subequations}
where $P_i=P_i(\mathbf{x},\mathbf{v},\mathbf{a})$ and $Q_i=Q_i(\mathbf{x},\mathbf{v},\mathbf{a})$ are scalar coefficient functions. This representation is complete, namely any other vector combination reducible via vector the triple product identity, $\mathbf{A}\times(\mathbf{B}\times\mathbf{C})=(\mathbf{A}\cdot\mathbf{C})\mathbf{B}-(\mathbf{A}\cdot\mathbf{B})\mathbf{C}$, collapses to these six basis terms \cite{Smith1971,Villar2023}.
Taking the time derivative of $\mathbf{p}$ in Eq. \eqref{MomentumDecomp_a} (l.h.s.) and comparing it term-by-term with the r.h.s. of Eq. \eqref{GTMomentum}, after invoking Eq. \eqref{MomentumDecomp_b}, we obtain
\begin{subequations}
\begin{eqnarray}
\nonumber
\dot{\mathbf{p}}_\mathrm{lhs}
&=&
\dot{P}_1\mathbf{x}+(P_1 + \dot{P}_2)\mathbf{v}+(P_2 + \dot{P}_3)\mathbf{a} +  P_3\dot{\mathbf{a}} \\
\nonumber
&+& 
\dot{P}_4(\mathbf{x}\times\mathbf{v})+ (P_4 + \dot{P}_5)(\mathbf{x}\times\mathbf{a}) + (P_5 + \dot{P}_6)(\mathbf{v}\times\mathbf{a})\\
&+& 
P_5(\mathbf{x}\times\dot{\mathbf{a}}) + P_6(\mathbf{v}\times\dot{\mathbf{a}})\,,\\
\dot{\mathbf{p}}_\mathrm{rhs}
\nonumber
&=&
[Q_4v^2 + Q_5(\mathbf{v}\cdot\mathbf{a})]\mathbf{x}
+
[-Q_4(\mathbf{x}\cdot\mathbf{v})+Q_6(\mathbf{v}\cdot\mathbf{a})]\mathbf{v}
+
[f(v) - Q_5(\mathbf{x}\cdot\mathbf{v})-Q_6v^2]\mathbf{a}\\
&-& 
Q_1(\mathbf{x}\times\mathbf{v})
+
Q_3(\mathbf{v}\times\mathbf{a})\,.
\end{eqnarray}
\end{subequations}
These two vectors are equal, i.e. $\dot{\mathbf{p}}_\mathrm{lhs}=\dot{\mathbf{p}}_\mathrm{rhs}$, if the coefficients of the respective state vectors are equal. Then, comparing them, we have
\begin{subequations}\label{Conditions}
\begin{eqnarray}
\label{Conditions_a}
\dot{P}_1 &=& Q_4v^2 + Q_5(\mathbf{v}\cdot\mathbf{a})\,,\\
\label{Conditions_b}
P_1 + \dot{P}_2 &=& -Q_4(\mathbf{x}\cdot\mathbf{v}) + Q_6(\mathbf{v}\cdot\mathbf{a})\,,\\
\label{Conditions_c}
P_2 + \dot{P}_3 &=& f(v) - Q_5(\mathbf{x}\cdot\mathbf{v})-Q_6v^2\,,\\
\label{Conditions_d}
P_3 &=& 0\,,\\
\dot{P}_4 &=& - Q_1\,,\\
P_4 + \dot{P}_5 &=& 0\,,\\
P_5 + \dot{P}_6 &=& Q_3\,,\\
\label{Conditions_h}
P_5 &=& 0\,,\\
\label{Conditions_i}
P_6 &=& 0\,.
\end{eqnarray}
\end{subequations}
From these, we immediately deduce from Eqs. \eqref{Conditions_d}-\eqref{Conditions_i}
\begin{eqnarray}\label{SecIIeq10}
P_3 = P_4=P_5=P_6=0\,,\qquad Q_1=Q_3=0\,.
\end{eqnarray}
Substituting back into Eqs. \eqref{Conditions_b}-\eqref{Conditions_c}, we find 
\begin{subequations}\label{SimiplConds}
\begin{eqnarray}
\label{SimiplConds_a}
P_1&=&-\dot{f}(v)+(\dot{Q}_5-Q_4)(\mathbf{x}\cdot\mathbf{v})+Q_5(\mathbf{x}\cdot\mathbf{a}) + 3Q_6(\mathbf{v}\cdot\mathbf{a})+(Q_5+\dot{Q}_6)v^2\,,\\
\label{SimiplConds_b}
P_2 &=&
f(v)-Q_5(\mathbf{x}\cdot\mathbf{v})-Q_6 v^2\,,
\end{eqnarray}
\end{subequations}
with arbitrary $Q_2=Q_2(\mathbf{x},\mathbf{v},\mathbf{a})$. Taking the time derivative of $P_1$ and enforcing Eq. \eqref{Conditions_a} leads to a crucial condition coupling $f(v)$ and $Q_{i=4,5,6}$,
\begin{eqnarray}\label{Constaint1}
\ddot{f}(v) 
=
\mathbf{x}\cdot 
\frac{\md}{\md t}
\left[ 
\left(\dot{Q}_5-Q_4\right)\mathbf{v}
+ 
Q_5\mathbf{a}
\right]
+
2Q_5(\mathbf{v}\cdot\mathbf{a})  
+ 
2\left(\dot{Q}_5-Q_4\right)v^2 
+
\frac{\md}{\md t} \left[Q_6 (\mathbf{v}\cdot\mathbf{a}) + \frac{\md}{\md t}(v^2 Q_6)\right]\,.
\end{eqnarray}
Since the l.h.s. of Eq. \eqref{Constaint1} depends only on the scalar speed $v$ and its time derivatives, consistency requires cancellation of all explicit dependence on the position vector $\mathbf{x}$ appearing on the r.h.s. The minimal way to enforce this, without introducing additional compensating constraints among independent kinematic variables, is to set $Q_4=Q_5=0$. The remaining coefficient $Q_6$ may then depend only on variables compatible with the scalar structure of the l.h.s., reducing to $Q_6(\mathbf{v},\mathbf{a})$.
%
%
%
%
%
Thus, Eq. \eqref{Constaint1} simplifies to
\begin{eqnarray}
\ddot{f}(v)
&=&
\frac{\md}{\md t} \Big[Q_6 (\mathbf{v}\cdot\mathbf{a}) + \frac{\md}{\md t}(v^2 Q_6)\Big]\,,
\end{eqnarray}
from which we can solve w.r.t. $Q_6$ as,
\begin{eqnarray}\label{ImpDiffEq}
Q_6(v)
&=&
\frac{1}{v^3}\left(\int v \frac{\partial f(v)}{\partial v}\md v - \alpha\right)\,,
\end{eqnarray}
where $\alpha$ is an integration constant. Here we observe two things, i. this specific  coupling between $f$ and $Q_6$ implies $P_1=0$, and ii. Eq. \eqref{ImpDiffEq} reveals that the dependence of the function $Q_6$ on the state variables $\{\mathbf{v},\mathbf{a}\}$ is reduced to $Q_6(\mathbf{v})$, and more precisely to $Q_6(v)$.

Collecting the results, we have
\begin{subequations}\label{MomExp}
\begin{eqnarray}
P_2(v)&=& \frac{1}{v}\left(\int f(v)\md v + \alpha  \right)\,,\\
Q_6(v)&=& \frac{1}{v^2}\Big( f(v) - P_2(v)\Big)\,,\\
\label{MomExp_a}
\mathbf{p}(\mathbf{v})&=& P_2(v)\mathbf{v}\,,\\
\label{MomExp_b}
\mathbf{M}(\mathbf{x},\mathbf{v},\mathbf{a}) &=&
Q_2(\mathbf{x},\mathbf{v},\mathbf{a})\mathbf{v} + Q_6(v)(\mathbf{v}\times\mathbf{a})\,,\\
\label{MomExp_c}
\mathbf{F}_\mathrm{c}(\mathbf{x}) + \mathbf{F}_{\perp}(\mathbf{x},\mathbf{v})
&=&
P_2(v)\mathbf{a} + Q_6(v)(\mathbf{v}\cdot\mathbf{a})\mathbf{v}
=
P_2(v)\mathbf{a}_{\perp} + f(v)\mathbf{a}_{\parallel}
\,,
\end{eqnarray}
\end{subequations}
where $\mathbf{a}_{\parallel}=\frac{(\mathbf{v}\cdot\mathbf{a})}{v^2}\mathbf{v}$ and $\mathbf{a}_{\perp}=\mathbf{a}-\mathbf{a}_{\parallel}$. Thus, once $f(v)$ is fixed, all the involved quantities in Eq. \eqref{MomExp} are determined. 
The $Q_2$-term multiplies a vector parallel to $\mathbf{v}$, and therefore contributes no transverse response, since
$\mathbf{v}\times(Q_2\mathbf{v})=\mathbf{0}$.

We note however, that a more general formulation, in which the condition used to eliminate the position dependence in Eq. \eqref{Constaint1} is relaxed, would allow for $P_1\neq0$ and for coefficients such as $P_2$ to depend explicitly on position. In such a case, the momentum vector $\mathbf{p}$ would acquire an additional component along the position vector $\mathbf{x}$, and more generally the terms $Q_4$ and $Q_5$ would no longer vanish. The analysis of this extended framework lies beyond the scope of the present work and is deferred to future investigation.

In summary, the condition $\dot{E}=0$ provides a unifying principle that organizes the set of admissible force laws, with the function $f(v)$ serving as the key bridge between kinetic energy and the system’s dynamical response. In its general form, the resulting force law is given in Eq.~\eqref{GNLaw}, while enforcing $SO(3)$-equivariance symmetry refines it to the more specific structure shown in Eq.~\eqref{MomExp_c}. This establishes a clear functional relation between kinetic energy and force, setting the stage for the subsequent determination of kinetic energy and momentum under IRF invariance.

\section{Relativity Principle and Invariance of the Inertial Response in One Dimension}\label{SecIII}
In this section we derive a purely kinematic result concerning the transformation of the total force in one-dimensional motion, to use it in the next section to determine the explicit expression of the response function $f(v)$. For simplicity we will omit the subscript ``total" in the force.

Consider two IRFs $\Sigma$ and $\Sigma'$, where $\Sigma'$ moves at constant velocity $u$ relative to $\Sigma$. In 1D motion the total force acting on the particle as measured in $\Sigma$ may be expressed as a scalar function of the kinematical quantities measured in that frame  $F=\Psi(v,a)$. Let $\Phi_u:\mathbb{R}\to\mathbb{R}$ denote the map that assigns to a force value $F$ measured in $\Sigma$ the corresponding force value $F'$ measured in $\Sigma'$, for a fixed relative velocity $u$. Then, RP requires that the dynamical law relating force to kinematics have the same functional form in all inertial frames. Accordingly, the transformed force must satisfy
\begin{eqnarray}
F'=\Phi_u(F)=\Phi_u(\Psi(v,a))=\Psi(v',a')\,,
\end{eqnarray}
where $v'$ and $a'$ are the velocity and acceleration of the particle as measured in $\Sigma'$.

To proceed, we emphasize that the Relativity Principle must be supplemented by a small set of standard structural assumptions concerning the transformation of forces, analogous to those commonly invoked in kinematic derivations of inertial-frame transformations. In particular, the force transformation maps $\Phi_u$ are required to satisfy physically natural properties such as additivity (superposition), monotonicity, isotropy, and continuity. These conditions do not introduce new dynamical content, but rather ensure that force transformations are regular, non-pathological, and compatible with the absence of preferred directions or force scales in inertial frames.

This strategy parallels the approach adopted in Ref. \cite{Berzi199}, who showed that, in 1D kinematics, the RP together with analogous structural assumptions, continuity, monotonicity, and reciprocity, suffices to uniquely constrain the admissible velocity transformation laws. Here we adopt the same methodological perspective, applying it at the level of force transformations rather than kinematics. Guided by this analogy, we impose the following requirements:
\begin{itemize}
\item[(i)] \textit{Additivity (superposition)}: Independent forces add linearly in every inertial frame, so
\[
\Phi_u(F_1+F_2)=\Phi_u(F_1)+\Phi_u(F_2) \qquad \text{for all }F_1,F_2\in\mathbb{R}\,.
\]
To understand this property, consider two forces $F_1$ and $F_2$ acting simultaneously on a particle in $\Sigma$. In $\Sigma'$, the total force is $F' = \Phi_u(F_1 + F_2)$.  
On the other hand, applying the RP to each force individually, we have $F_1' = \Phi_u(F_1)$ and $ F_2' = \Phi_u(F_2)$.
Since the forces are independent and superpose linearly, the total force in $\Sigma'$ must satisfy
$F' = F_1' + F_2' = \Phi_u(F_1) + \Phi_u(F_2)$, which precisely gives the additivity condition for the transformation $\Phi_u$.
\item[(ii)] \textit{Order preservation (monotonicity)}:  If $F_1>F_2$ in $\Sigma$, then in $\Sigma'$ we should have $F_1'>F_2'$. This ensures that stronger forces remain stronger and repulsive/attractive directions are not flipped by the boost. Mathematically, this means that for a fixed $u$, the map $\Phi_u$ must satisfy
\[ 
F_1>F_2\quad\Rightarrow\quad \Phi_u(F_1)>\Phi_u(F_2)\,.
\]
Equivalently, $\Phi_u$ is a monotone increasing function of $F$.
\item[(iii)] \textit{Inverse--boost property}: If we boost from $\Sigma$ to $\Sigma'$ with velocity $u$, and then return from $\Sigma'$ to $\Sigma$ with velocity $-u$, the force measured in the original frame must be recovered. Mathematically, this requires
\[
\Phi_{-u}\big(\Phi_{u}(F)\big) = F \qquad \text{for all } F \in \mathbb{R}.
\]
This property ensures that the transformation is invertible, with the inverse of $\Phi_{u}$ given by $\Phi_{-u}$.
\item[(iv)] \textit{Identity at zero relative velocity}: When there is no relative motion between frames, the force measured in $\Sigma'$ must coincide with that in $\Sigma$. Mathematically, this reads
\[
\Phi_0(F) = F \qquad \text{for all } F \in \mathbb{R}\,.
\]
This reflects the trivial case of a zero-velocity boost, where no transformation occurs.
\item[(v)] \textit{Continuity (or smoothness) in the boost parameter}: The force measured in $\Sigma'$ should vary continuously with the relative velocity $u$. Therefore, for each fixed $F$, the map $u \mapsto \Phi_{u}(F)$ is assumed to be a smooth function of $u$. This ensures that small changes in the relative velocity produce small changes in the measured force, consistent with physical intuition.
\end{itemize}
In Appendix \ref{AppendixA} we show that, taken together, the above structural properties of 
$\Phi_u$ uniquely imply that the force is invariant under one-dimensional boosts,
\begin{eqnarray}\label{Invar0000}
F' = F \,,
\end{eqnarray}
and therefore substituting $F$ and $F'$ from Eq. \eqref{GTMomentum} we have
\begin{eqnarray}\label{Invar00}
f(v)a = f(v')a' \,,
\end{eqnarray}
for any two IRFs related by a 1D relative motion. The second relation follows from the fact that, in one dimension, the transverse term $\mathbf{v}\times\mathbf{M}$ vanishes identically. We refer to Eqs.~\eqref{Invar0000} and \eqref{Invar00} as the \textit{1D Force-Boost Invariance} (1D-FBI).  
In other words, in strictly 1D motion the RP, together with the above structural constraints on the force transformation, entails that the longitudinal force is unchanged by inertial boosts:
\begin{eqnarray*}
\text{Relativity Principle } + \text{ Structural Constraints } \Rightarrow \text{ 1D Force-Boost Invariance.}
\end{eqnarray*}

The FBI result obtained in one dimension relies crucially on the scalar character of the transformed quantity, which allows the above requirements to act jointly and thereby fix the transformation uniquely. In higher dimensions, the force is a vector and the corresponding transformation laws are vector-valued maps. In this setting, the same physical assumptions do not uniquely lead to $\mathbf{F}'=\mathbf{F}$. However, since our goal is to determine $f(v)$ for a given coordinate transformation, the aforementioned limitation does not affect the generality of Eq.~\eqref{Invar00}, as $f(v)$ is a scalar function whose form is independent of the dimensionality of the motion.

\section{Determination of $f(v)$ from 1D-FBI under Galilean and Lorentz Kinematics}\label{SecIIIb}

In this section we make use of the 1D-FBI to determine the explicit form of the velocity–response function $f(v)$. The only additional input now required is the kinematical relation between the velocities and accelerations measured in two inertial reference frames in relative motion.
Galilean and Lorentz transformations will therefore enter here as coordinate relations between inertial frames. Their forms may be obtained from general symmetry principles, linearity, homogeneity, isotropy, and reciprocity, without appeal to any particular spacetime geometry or dynamical assumptions~\cite{JeanMarc1976}. Once these kinematical relations are specified, the functional form of $f(v)$ follows directly from 1D–FBI.

\subsection{Galilean Transformations}

The Galilean 1D-transformation law between two IRFs $\Sigma$ and $\Sigma'$ with the relative speed $u_0$ is given by
\begin{equation}
t'=t\,,
\qquad
x'=x-u_0 t, 
\qquad 
v'=v-u_0,
\qquad 
a'=a.
\end{equation}
Substituting these relations into Eq.~\eqref{Invar00}, for $a\neq0$, we obtain
\begin{equation}
f(v)=f(v-u_0)\qquad \forall v,u_0\in\mathbb{R}.
\end{equation}
Setting $v=u_0$ gives $f(u_0)=f(0)$, and since $u_0$ is arbitrary, $f(v)$ must be constant, thus
\begin{equation}\label{ffunc}
f(v)=m=const.\,.
\end{equation}
Substituting Eq. \eqref{ffunc} into Eq. \eqref{MomExp_a}, the momentum vector becomes $\mathbf{p}(\mathbf{v}) = (mv + \alpha) \hat{\mathbf{v}}$, where $\hat{\mathbf{v}}$ is the velocity unit vector. The momentum vector must be a well-defined vector-valued function for every admissible velocity $\mathbf{v}$. In the inertial state $\mathbf{v}=\mathbf{0}$, the direction $\hat{\mathbf{v}}$ is undefined. If $\alpha\neq0$,  $\mathbf{p}(\mathbf{0})$ would be
ill-defined. To ensure that momentum is well defined for all velocities, including $\mathbf{v}=\mathbf{0}$, we must therefore set 
$\alpha=0$, yielding $\mathbf{p}(\mathbf{0})=\mathbf{0}$.

With $f(v)$ fixed, the kinetic energy, momentum, and force law follow directly from the definitions in Eq.~\eqref{MomExp}. Thus, for Galilean-invariant dynamics, these quantities take the form
\begin{subequations}\label{GalKinEnALL}
\begin{eqnarray}
\label{GalKinEn}
T(v)&=&\frac{1}{2} m v^2 + d_1\,,\\
\mathbf{p}(\mathbf{v})&=& m\mathbf{v}\,,\\
\mathbf{F}_\mathrm{tot}
&=&
m\mathbf{a}\,,
\end{eqnarray}
\end{subequations}
where $d_1$ is an integration constant. Setting it to zero, $d_1=0$, the former expressions recover the standard Newtonian expressions for kinetic energy, momentum, and force, while $m$ is identified with the mass of the particle. In this way, the familiar structure of Newtonian mechanics appears as a specific realization of the more general energy-based framework, selected by Galilean kinematics.

It is worth emphasizing that, for Galilean transformations, the force is invariant between IRFs. Indeed, since the acceleration is the same in all Galilean-related frames and the inertial response is fixed to $f(v)=m$, the total force satisfies $\mathbf{F}'=\mathbf{F}$.

\subsection{Lorentz Transformations}\label{secIVc}

Having shown how the Galilean forms of $T$ and $\mathbf{p}$ follow solely from 1D-FBI in Eq. \eqref{Invar00}, we now extend the argument to Lorentz 1D-transformations for a motion along the $x$-axis with the relative speed $u_0$, which are given as
\begin{eqnarray} \label{LT1} 
t' = \gamma(u_0)\left(t- \frac{u_0 x}{c^2} \right)\,,\hskip1.0cm 
x' = \gamma(u_0) (x-u_0 t)\,,\hskip1.0cm 
v' = \frac{v-u_0}{1-\ds\frac{u_0v}{c^2}}\,, \hskip0.95cm 
a' = \frac{a}{\gamma^3(u_0)\left(1-\ds\frac{u_0 v}{c^2}\right)^3}\,,
\end{eqnarray} 
with $\gamma(v):=(1-v^2/c^2)^{-1/2}$. Notice that, also in this relativistic setting, time $t$ is treated as an external evolution parameter associated with each inertial frame. The Lorentz transformation is employed solely as a kinematic relation between the coordinates $(x,x')$ of different inertial frames, without introducing a spacetime metric or promoting time to an additional geometric dimension. The following steps are conceptually the same as in the Galilean case above.

Substituting Eq.~\eqref{LT1} into Eq.~\eqref{Invar00}, with   $a\neq 0$, gives
\begin{equation}\label{fLT}
f(v)=
f\left(\frac{v-u_0}{1-\ds\frac{u_0v}{c^2}}\right)
\frac{1}{\gamma^3(u_0)\left(1-\ds\frac{u_0 v}{c^2}\right)^3}.
\end{equation}
Setting as in the Galilean case $v=u_0$ yields $f(u_0)=f(0)\gamma^3(u_0)$. Since $u_0$ is arbitrary and for $f(0)=\mu\neq0$, we obtain
\begin{equation}\label{fLT2}
f(v)=\mu\left(1-\frac{v^2}{c^2}\right)^{-3/2}.
\end{equation}
Using Eq.~\eqref{fdef} then gives
\begin{equation}
T(v) = \frac{\mu c^2}{\ds\sqrt{1-\frac{v^2}{c^2}}} + d_2.
\end{equation}
Requiring the Newtonian limit as $c\to\infty$ identifies $\mu=m$ and $d_2=-mc^2+d_1$, so
\begin{equation}\label{relKinEn}
T(v) = mc^2\left(\frac{1}{\ds\sqrt{1-\frac{v^2}{c^2}}}-1\right) + d_1 ,
\end{equation}
the standard relativistic kinetic energy.

Substituting Eq.~\eqref{fLT2} into Eqs.~\eqref{MomExp}, setting again $d_1=0$ and $\alpha=0$ for the same reason presented in the Galilean case, yields
\begin{subequations}
\begin{eqnarray}
T(v) &=& mc^2\big(\gamma(v)-1\big)\,,\\
\label{RelMom}
\mathbf{p}(\mathbf{v})
&=&
m\gamma(v)\mathbf{v}\,,\\
\label{RelForceLaw}
\mathbf{F}_\mathrm{tot}
&=&
m\gamma^3(v)\left[\mathbf{a} + \frac{1}{c^2}\mathbf{v}\times(\mathbf{v}\times \mathbf{a}) \right]
=
m\gamma(v)\left[\mathbf{a} + \frac{\gamma^2(v)}{c^2}(\mathbf{v}\cdot\mathbf{a})\mathbf{v} \right]
=
m\gamma(v) \mathbf{a}_{\perp}+m\gamma^3(v)\mathbf{a}_{\parallel}\,.
\end{eqnarray}
\end{subequations}
These expressions correspond to the standard results of special relativity, including Planck’s relativistic momentum~\cite{Planck1906}.

The term proportional to $\mathbf{v}\times\mathbf{a}$ arising from the $\mathbf{M}$ vector is therefore not optional: 1D-FBI uniquely determines its coefficient. Whenever the kinetic energy deviates from the quadratic Newtonian form, a transverse inertial response necessarily appears, reflecting the experimentally observed anisotropy of relativistic dynamics.

Finally, we note that although the conventional formulation of special relativity is expressed geometrically in terms of Minkowski spacetime~\cite{Catoni2011,Naber2012}, the derivation presented here shows that the same relativistic dynamical structure can be obtained directly from local energy conservation combined with 1D–FBI, without explicitly invoking spacetime geometry. Nevertheless, when expressed as a 3D force law, the resulting relativistic force is neither Lorentz invariant nor Lorentz covariant. The latter means that the quantity 
$\md \mathbf{p}/\md t$ does not transform as a tensorial object under Lorentz transformations, and the corresponding force law acquires a frame-dependent form. As is well known, maintaining covariance of the equations of motion under Lorentz boosts for general 3D motion then requires an extension of the framework, most commonly achieved by embedding the theory in 4D Minkowski spacetime, where space and time are unified into a single geometric structure.

\begin{figure}[t]
\centering
\resizebox{0.8\linewidth}{!}{%
\begin{tikzpicture}[
    node distance=8mm and 10mm,
    rounded corners=2pt,
    every node/.style={align=center, font=\small},
    block/.style={draw, thick, rounded corners, fill=gray!10,
                  text width=4.4cm, inner sep=4pt},
    topblock/.style={draw, thick, rounded corners,
                     fill=red!15, text width=5.0cm, inner sep=5pt},
    titleblock/.style={draw, thick, rounded corners,
                       fill=blue!15, text width=5.0cm, inner sep=5pt},
    titleblockSO/.style={draw, thick, rounded corners,
                       fill=blue!15, text width=5.0cm, inner sep=5pt},
    highlightblock/.style={draw, thick, rounded corners,
                           fill=green!15, text width=4.3cm, inner sep=4pt},
    arrow/.style={-{Latex[length=2mm]}, thick}
]

\node[topblock] (energy)
{(Local energy balance: PCE) \\ $\displaystyle \dot{E} \!\!\!=\!\!\! 0$ 
};

\node[titleblock, below=10mm of energy] (genlaw)
{$\displaystyle 
\mathbf{F}_{\mathrm{tot}}
= f(v)\,\mathbf{a} \;+\; \mathbf{v}\!\times\!\mathbf{M}$};

\draw[arrow] (energy) -- (genlaw);

\node[titleblockSO, below=10mm of genlaw] (genlawSO)
{$SO(3)$-Equivariance \\ $\displaystyle \mathbf{F}_{\mathrm{tot}}
= f(v)\mathbf{a}_{\parallel}+ \left(\frac{1}{v}\int f(v)\md v\right) \mathbf{a}_{\perp}$};

\draw[arrow] (genlaw) -- (genlawSO);

\node[highlightblock, below=10mm of genlawSO, text width=6.0cm] (principle)
{
1D Relativity + Structural Constraints \\[1mm]
$\Rightarrow$\hskip0.5cm \boxed{%
    \parbox{2.5cm}{\centering
    1D-FBI: $F_\mathrm{tot}'=F_\mathrm{tot}$
}}
};

\draw[arrow] (genlawSO) -- (principle);

\node[block, below left=10mm and -2mm of principle]
(classical)
{Galilean Transformation};

\node[block, below right=10mm and -2mm of principle]
(relativistic)
{Lorentz Transformation};

\draw[arrow] (principle) -- (classical);
\draw[arrow] (principle) -- (relativistic);

\node[block, below=6mm of classical]
(classical2)
{\vskip-0.7cm\begin{eqnarray*}
f(v) = m   \quad \Rightarrow\; 
\begin{cases}
 T = \frac{1}{2} m v^2\\
\mathbf{p} = m \mathbf{v}\\
\mathbf{v}\!\times\!\mathbf{M} = \mathbf{0}
\end{cases}
\end{eqnarray*}
};

\draw[arrow] (classical) -- (classical2);

\node[block, below=6mm of classical2]
(classical3)
{\vskip-0.1cm 3D-FBI: \hskip1.0cm $\mathbf{F}'_\mathrm{tot}=\mathbf{F}_\mathrm{tot}$
};

\draw[arrow] (classical2) -- (classical3);

\node[block, below=6mm of relativistic, text width=7.3cm]
(relativistic2)
{\vskip-0.7cm\begin{eqnarray*}
f(v) = m \gamma^3(v)  \quad \Rightarrow\;
\begin{cases}
 T = m c^2 \big(\gamma(v) - 1\big)\\
\mathbf{p} = m \gamma(v) \mathbf{v}\\
\mathbf{v}\!\times\!\mathbf{M} = \frac{m}{c^2} \gamma^3(v)\, \mathbf{v}\!\times\!(\mathbf{v}\!\times\!\mathbf{a})   
\end{cases}
\end{eqnarray*}
};

\draw[arrow] (relativistic) -- (relativistic2);

\node[block, below=6mm of relativistic2, text width=7.3cm]
(relativistic3)
{\vskip-0.1cm 3D-FBI:\hskip1.0cm $\widetilde{\mathbf{F}}'_\mathrm{tot}=\widetilde{\mathbf{F}}_\mathrm{tot}$
};

\draw[arrow] (relativistic2) -- (relativistic3);


\end{tikzpicture}%
}
\caption{Conceptual Roadmap: Newtonian and relativistic dynamics emerge from local energy balance and 1D Force-Boost Invariance.}
\label{fig:roadmap}
\end{figure}
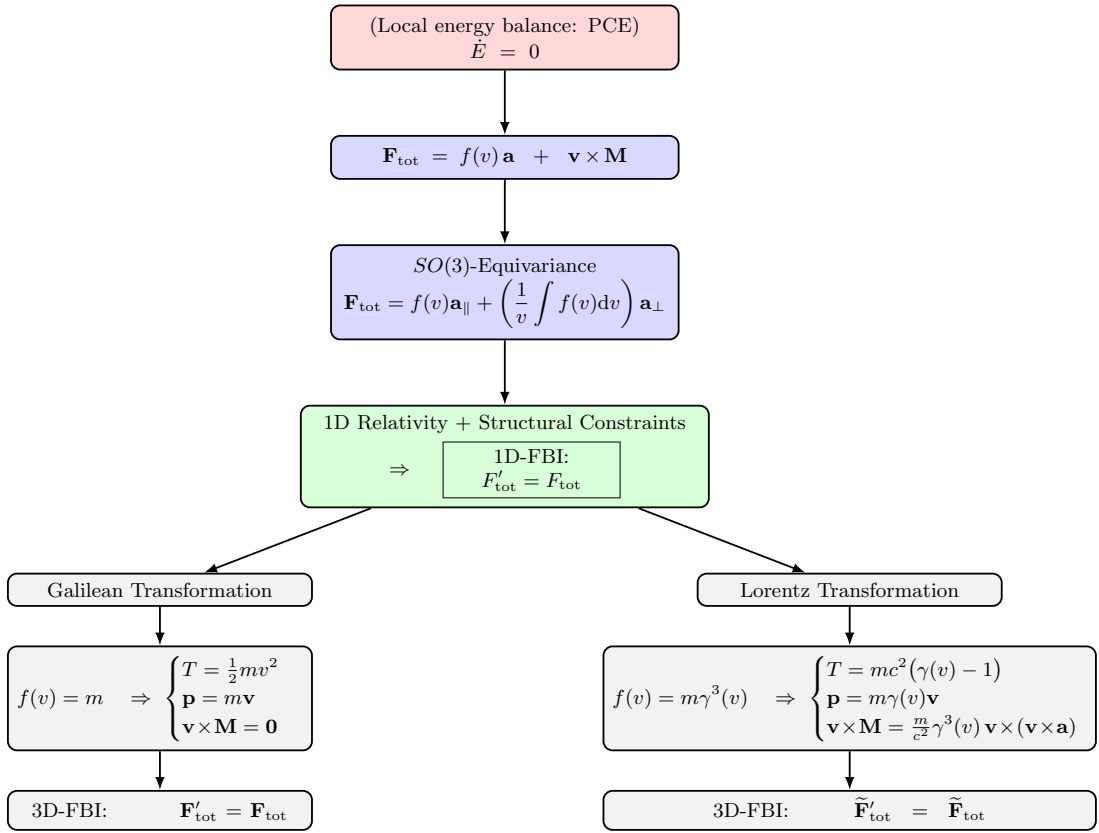

An alternative perspective has been proposed by R\c ebilas \cite{Rebilas2010} within a purely Euclidean framework. In particular, he demonstrated that the vector $\left(\frac{\md \mathbf{p}}{\md t}\right)_{\parallel} + \gamma(v)\left(\frac{\md \mathbf{p}}{\md t}\right)_\perp$ is invariant in all IRFs and coincides with the total force in the instantaneous rest frame of the particle. This construction restores a form of frame-independent description within a purely three-dimensional setting, providing an invariant characterization of the dynamical response without recourse to four-vectors. Rearranging the terms in the relativistic force law in Eq.~\eqref{RelForceLaw} as
\begin{eqnarray}
\mathbf{F}_\mathrm{tot} + m\gamma(v)\big(\gamma(v)-1\big)\mathbf{a}_\perp= m\gamma^2(v)\mathbf{a}_\perp + m\gamma^3(v)\mathbf{a}_\parallel\,,
\end{eqnarray}
we identify the r.h.s. with this invariant combination. This motivates the definition of a new momentum-like vector $\widetilde{\mathbf{p}}$ and an effective force vector $\widetilde{\mathbf{F}}_\mathrm{tot}$ through
\begin{subequations}
\begin{eqnarray}
\frac{\md \widetilde{\mathbf{p}}}{\md t}
&:=&
\left(\frac{\md \mathbf{p}}{\md t}\right)_{\parallel} + \gamma(v)\left(\frac{\md \mathbf{p}}{\md t}\right)_\perp
=
m\gamma^3(v)\mathbf{a}_\parallel + m\gamma^2(v)\mathbf{a}_\perp\,,\\
\widetilde{\mathbf{F}}_\mathrm{tot}
&:=&
\mathbf{F}_\mathrm{tot} + m\gamma(v)\big(\gamma(v)-1\big)\mathbf{a}_\perp
=
\mathbf{F}_\mathrm{tot} - \frac{1}{\gamma(v)+1}\mathbf{v}\times\mathbf{M}\,.
\end{eqnarray}
\end{subequations}
In terms of these quantities, the relativistic force law can be reformulated in the manifestly boost-invariant form
\begin{eqnarray}
\frac{\md \widetilde{\mathbf{p}} }{\md t}
=
\widetilde{\mathbf{F}}_\mathrm{tot}(\mathbf{x},\mathbf{v},\mathbf{a}) 
=
-\boldsymbol{\nabla}_\mathbf{x} V(\mathbf{x})
+
\mathbf{v}\times\widetilde{\mathbf{K}}(\mathbf{x},\mathbf{v},\mathbf{a})\,,\qquad
\widetilde{\mathbf{K}}(\mathbf{x},\mathbf{v},\mathbf{a})
:=
\mathbf{K}(\mathbf{x},\mathbf{v})
- 
\frac{1}{\gamma(v)+1}\mathbf{M}(\mathbf{v},\mathbf{a})\,,
 \end{eqnarray}
which correctly reduces to Newton’s second law in the nonrelativistic limit $\gamma\to1$, i.e., $\frac{\md \mathbf{p}}{\md t}=\mathbf{F}_\mathrm{tot}(\mathbf{x},\mathbf{v})$. In Fig. \ref{fig:roadmap} we present the conceptual roadmap of our results so far.

It is worth mentioning however, that R\c ebilas' approach does not yield a covariant formulation in the standard sense. The invariant quantity is not part of a four-vector or tensorial structure, and its construction relies on a decomposition relative to the particle velocity rather than on spacetime geometry. Consequently, while it provides an invariant reformulation of the force concept, it does not replace the fully Lorentz-covariant description based on four-momentum and proper time in Minkowski spacetime. The relation between such invariant three-dimensional formulations and the standard covariant framework therefore remains structurally nontrivial.

\section{Connection between PCE and PLA}\label{SecIV}
%

As emphasized in the Introduction, the PCE and PLA offer complementary perspectives on dynamical structure. The PLA provides a universal and powerful framework, particularly in field theory, yet the functional form of the Lagrangian is not uniquely determined; symmetry arguments can restrict possibilities but do not, by themselves, guarantee physical relevance. The PCE, in contrast, centers dynamics directly on kinetic-energy change and its transformation properties. In this section we make the connection between the two approaches explicit and identify the conditions under which their formulations coincide.

Consider the action functional
\begin{eqnarray}
S=\int_{t_1}^{t_2}L(\mathbf{x}(t),\mathbf{v}(t),t)\md t\,.
\end{eqnarray}
The Euler–Lagrange equations follow from extremization,
\begin{eqnarray}\label{ELEq}
\frac{\md}{\md t} \boldsymbol{\nabla}_{\mathbf{v}}L
=
\boldsymbol{\nabla}_{\mathbf{x}}L\,,
\end{eqnarray}
with $\boldsymbol{\nabla}_\mathbf{v}:=(\frac{\partial}{\partial v_1},\frac{\partial}{\partial v_2},\frac{\partial}{\partial v_3})^T$. 
A standard identity derived from Eq.~\eqref{ELEq} is
\begin{eqnarray}
    \frac{\md L}{\md t}
    =
    \mathbf{v}\cdot\boldsymbol{\nabla}_{\mathbf{x}}L
    +
    \mathbf{a}\cdot \boldsymbol{\nabla}_{\mathbf{v}}L
    +
    \frac{\partial L}{\partial t}
    \overset{(\ref{ELEq})}{=}
    \mathbf{v}\cdot\frac{\md }{\md t} \boldsymbol{\nabla}_{\mathbf{v}}L
    +
    \mathbf{a}\cdot \boldsymbol{\nabla}_{\mathbf{v}}L
    +
    \frac{\partial L}{\partial t}
    =
    \frac{\md }{\md t}\left(\mathbf{v}\cdot \boldsymbol{\nabla}_{\mathbf{v}}L \right)
    +
    \frac{\partial L}{\partial t}\,.
\end{eqnarray}
Re-ordering the terms we have
\begin{eqnarray}
    \frac{\md }{\md t}\left(\mathbf{v}\cdot \boldsymbol{\nabla}_{\mathbf{v}}L -L\right)
    =
    -\frac{\partial L}{\partial t}\,.
\end{eqnarray}
Thus, if $L$ has no explicit time dependence, the function
\begin{eqnarray}\label{HamFun1}
H(\mathbf{x},\mathbf{v})
:=
\mathbf{v}\cdot \boldsymbol{\nabla}_{\mathbf{v}}L(\mathbf{x},\mathbf{v})-L(\mathbf{x},\mathbf{v})
\end{eqnarray}
is conserved in time. Both $L$ and $H$ have the dimensions of energy.

Within PCE, the total mechanical energy is explicitly defined as $E=T+V$. To align PLA with this definition, we impose the physical identification $H \equiv E$, ensuring that the variationally derived invariant corresponds to the physically meaningful energy. 
This identification relies on the structural assumptions underlying the definition of mechanical energy in Eq. \eqref{eqII4}. In particular, it presupposes that the energy is separable as $E=T(v)+V(\mathbf{x})$, with no explicit time dependence,  no higher-order velocity–position couplings and no dissipation. Under these conditions, the former identification acts as a constraint on the admissible Lagrangians, taking the form
\begin{eqnarray}\label{EAAA1}
\mathbf{v}\cdot \boldsymbol{\nabla}_{\mathbf{v}}L(\mathbf{x},\mathbf{v}) - L(\mathbf{x},\mathbf{v})
=
T(v)+V(\mathbf{x})\,,
\end{eqnarray}
which  integrates to
\begin{eqnarray}\label{ExplLang}
L(\mathbf{x},\mathbf{v}) = G(v) + \mathbf{v} \cdot \mathbf{A}(\mathbf{x}) - V(\mathbf{x})\,, 
\qquad
G(v)
:=
v \int \frac{T(v)}{v^2} \md v\,.
\end{eqnarray}
The term $\mathbf{v} \cdot \mathbf{A}(\mathbf{x})$ is the homogeneous solution and the term $G(v) -V(\mathbf{x})$ is a particular solution of Eq.~\eqref{EAAA1}. Here $G(v)$ encodes the purely velocity-dependent (``kinetic’’) part of the Lagrangian, and $\mathbf{A}(\mathbf{x})$ accounts for velocity–position couplings.
A key point is that $G(v)$ is \textit{not} the kinetic energy but a specific functional of it. Differentiating twice, we have
\begin{eqnarray}\label{fANDG}
\frac{\partial^2}{\partial v^2}\left(\frac{G(v)}{v}\right)
=\frac{\partial}{\partial v}\left(\frac{T(v)}{v^2}\right)
\qquad\xLongrightarrow{\eqref{fdef}}\qquad
f(v)=\frac{\partial^2G(v)}{\partial v^2}\,,
\end{eqnarray}
so the functions $G$ compatible with PLA–PCE consistency are fully determined once $f(v)$ is fixed by inertial-frame invariance. Integrating the Galilean and Lorentzian forms of $f(v)$ from Sec.~\ref{SecIII} gives
\begin{eqnarray}
G_\mtn{\mathrm{G}}(v)=\frac{1}{2}mv^2+C_{\mtn{\mathrm{G}},1} v+C_{\mtn{\mathrm{G}},2} \qquad\text{and}\qquad 
G_\mtn{\mathrm{L}}(v)=-mc^2\sqrt{1-\frac{v^2}{c^2}}+C_{\mtn{\mathrm{L}},1} v+C_{\mtn{\mathrm{L}},2}\,,
\end{eqnarray}
where $C_{\mtn{\mathrm{G}},i}$, $C_{\mtn{\mathrm{L}},i}$ are integration constants, and the subscripts G and L refer to the Galilean and Lorentz-invariant cases, respectively.
For vanishing integration constants, $G_{\mtn{\mathrm{L}}}$ reduces to the kinetic term of the standard relativistic Lagrangian. Remarkably, this conclusion is reached without appealing to the Minkowski metric or any spacetime-geometric structure: it follows solely from local energy balance together with 1D–FBI, all within Euclidean space.

More importantly, the present analysis shows that the usual definition of the Lagrangian as $L=T-V$ implicitly enforces $G=T$, and therefore $H\equiv E$ with $T\propto v^{2}$. In other words, the conventional form $L=T-V$ inherently presupposes Galilean relativity.

To complete the PLA-PCE connection analysis, next step is to explore the force law of PLA. Comparing Eq. (\ref{ELEq}) with the momentum form of the force law, i.e., $\mathbf{F}_\mathrm{tot}=\md \mathbf{p}/\md t$, one can analogously define canonical momentum and canonical forces in terms of the Lagrangian function as
\begin{eqnarray}\label{pFcan1}
\mathbf{p}^{\mtn{\mathrm{can}}}
=
\boldsymbol{\nabla}_{\mathbf{v}}L
\overset{(\ref{ExplLang})}{=}
\boldsymbol{\nabla}_{\mathbf{v}}
(G + \mathbf{v} \cdot \mathbf{A} )\,,\qquad
\mathbf{F}^{\mtn{\mathrm{can}}}_\mathrm{tot}
=
\boldsymbol{\nabla}_{\mathbf{x}}L
\overset{(\ref{ExplLang})}{=}
-\boldsymbol{\nabla}_{\mathbf{x}}
\left(V
-
\mathbf{v} \cdot \mathbf{A} \right)\,.
\end{eqnarray}
It is important to recognize that, in general, the canonical momentum $\mathbf{p}^{\mtn{\mathrm{can}}}$ and canonical force $\mathbf{F}^{\mtn{\mathrm{can}}}_\mathrm{tot}$, derived from the Lagrangian, do not coincide with the momentum $\mathbf{p}$ and total force $\mathbf{F}_\mathrm{tot}$ obtained from the energy-based formulation. This distinction becomes clear upon comparing Eqs. (\ref{GNLaw}), (\ref{GTMomentum}), and (\ref{pFcan1}).
Bridging this gap offers valuable insight into the structure of dynamics, namely achieving full equivalence between the two formulations requires specific conditions that link the underlying functional forms. To make this connection explicit, we begin by considering the following vector identity for a vector $\mathbf{A}=\mathbf{A}(\mathbf{x})$, and a nabla operator that may operate either on $\mathbf{x}$ or on $\mathbf{v}$,
\begin{eqnarray}
\boldsymbol{\nabla}(\mathbf{v}\cdot\mathbf{A})
=
(\mathbf{v}\cdot\boldsymbol{\nabla})\mathbf{A}
+
(\mathbf{A}\cdot \boldsymbol{\nabla})\mathbf{v}
+
\mathbf{v}\times(\boldsymbol{\nabla}\times\mathbf{A})
+
\mathbf{A}\times(\boldsymbol{\nabla}\times\mathbf{v})\,.
\end{eqnarray}
Specifically, for $\boldsymbol{\nabla}=\boldsymbol{\nabla}_\mathbf{x}$ and $\boldsymbol{\nabla}=\boldsymbol{\nabla}_\mathbf{v}$, the former vector identity reduces to
\begin{subequations}\label{calc1}
\begin{eqnarray}
\label{calc1_b}
\boldsymbol{\nabla}_{\mathbf{v}}
(\mathbf{v}\cdot\mathbf{A})
&=&
\mathbf{A}\,,\\
\label{calc1_a}
\boldsymbol{\nabla}_{\mathbf{x}}(\mathbf{v}\cdot\mathbf{A})
&=&
\left(\mathbf{v}\cdot \boldsymbol{\nabla}_{\mathbf{x}}\right)\mathbf{A}
+
\mathbf{v}\times\left(\boldsymbol{\nabla}_{\mathbf{x}}\times\mathbf{A}\right) \,,
\end{eqnarray}
\end{subequations}
respectively, so that
\begin{eqnarray}\label{calc1_c}
\frac{\md }{\md t}\boldsymbol{\nabla}_{\mathbf{v}}
(\mathbf{v}\cdot\mathbf{A})
\overset{\text{(\ref{calc1_b})}}{=}
\frac{\md \mathbf{A}}{\md t}
=
\left(\mathbf{v}\cdot\boldsymbol{\nabla}_{\mathbf{x}}\right)\mathbf{A}
\overset{\text{(\ref{calc1_a})}}{=}
\boldsymbol{\nabla}_{\mathbf{x}}(\mathbf{v}\cdot\mathbf{A})
-
\mathbf{v}\times\left(\boldsymbol{\nabla}_{\mathbf{x}}\times\mathbf{A}\right)\,.
\end{eqnarray}
Then, substituting the Lagrange function in Eq. (\ref{ExplLang}) into the Euler-Lagrange equations in Eq. (\ref{ELEq}), we have
\begin{eqnarray}
\frac{\md}{\md t} \boldsymbol{\nabla}_{\mathbf{v}}[G(v) + \mathbf{v} \cdot \mathbf{A}(\mathbf{x}) - V(\mathbf{x})]
&=&
\boldsymbol{\nabla}_{\mathbf{x}}[G(v) + \mathbf{v} \cdot \mathbf{A}(\mathbf{x}) - V(\mathbf{x})]\,,
\end{eqnarray}
and invoking Eq. \eqref{calc1_c} we obtain
\begin{eqnarray}\label{ELEq3}
\frac{\md}{\md t} 
\boldsymbol{\nabla}_{\mathbf{v}}G(v)
=
- 
\boldsymbol{\nabla}_{\mathbf{x}}V(\mathbf{x})
+
\mathbf{v}\times[\boldsymbol{\nabla}_{\mathbf{x}}\times\mathbf{A}(\mathbf{x})]\,.
\end{eqnarray}
By comparing Eq. (\ref{ELEq3}) with Eqs. (\ref{GNLaw})–(\ref{GTMomentum}), we identify the equivalence conditions between PCE and PLA regarding the equations of motion,
\begin{subequations}
\begin{eqnarray}
\label{CrossProdM0}
\mathbf{p}
&=&
\boldsymbol{\nabla}_{\mathbf{v}} G(v)\,,\\
\label{CrossProdM1}
f(v)\mathbf{a}+\mathbf{v}\times\mathbf{M}
&=&
\frac{\md}{\md t}
\boldsymbol{\nabla}_{\mathbf{v}} G(v)\,,\\
\label{CrossProdM2}
\mathbf{K}
&=&
\boldsymbol{\nabla}_{\mathbf{x}}\times\mathbf{A}(\mathbf{x})\,.
\end{eqnarray}
\end{subequations}
After some simple algebra applied to Eq.~(\ref{CrossProdM1}), in which we express $f$ in terms of $G$, as in Eq.~\eqref{fANDG}, and compute the time derivative on the r.h.s. (see Appendix \ref{AppendixB}), we can rewrite the equation as
\begin{eqnarray}\label{eq41}
	\mathbf{v}\times\mathbf{M}(\mathbf{x},\mathbf{v},\mathbf{a})
	=
	\frac{1}{v^2}\left(\frac{\partial^2 G(v)}{\partial v^2}-\frac{1}{v}\frac{\partial G(v)}{\partial v}\right)\mathbf{v}\times(\mathbf{v}\times\mathbf{a})
	=
	Q_6(v)\,\mathbf{v}\times(\mathbf{v}\times\mathbf{a})\,.
\end{eqnarray}
Re-expressing $G$ in terms of $f$, we identify the scalar coefficient on the r.h.s. with $Q_6$. This result is consistent with the general form of $\mathbf{M}$ in Eq.~\eqref{MomExp_b}, obtained under the conditions $Q_1=Q_3=Q_4=Q_5=0$. In particular, the term proportional to $Q_2\mathbf{v}$ does not contribute here, since $\mathbf{v}\times(Q_2\mathbf{v})=\mathbf{0}$, and therefore the only surviving transverse structure is necessarily proportional to $\mathbf{v}\times(\mathbf{v}\times\mathbf{a})$.

The PCE and PLA formulations therefore coincide provided the Lagrangian takes the form in Eq.~\eqref{ExplLang}, $G(v)$ is determined by the frame-invariance condition in Eq. \eqref{Invar00}, and the transverse inertial response is encoded by Eq.~\eqref{eq41}. More general choices of the PCE coefficient functions $Q_i$ remain beyond the reach of the standard PLA expression~\eqref{ExplLang}. In particular, Eq.~\eqref{CrossProdM2} restricts $\mathbf{K}$ to depend only on position, whereas PCE allows dependence on both position and velocity. Therefore, PCE defines a broader class of admissible dynamics. The usual Lagrangian formulation appears as a special case in which the energy-based structure enforces a compatible variational form. This places PCE as a natural, symmetry-consistent foundation for both classical and relativistic particle mechanics, while opening paths for generalizations that the traditional PLA does not natively accommodate. An  illustration of this broader class is provided by non-holonomic mechanical systems with rolling constraints. Such systems conserve mechanical energy and exhibit velocity-dependent reaction forces orthogonal to the motion, consistent with the structure $\mathbf{F}_\perp=\mathbf{v}\times\mathbf{K}(\mathbf{x},\mathbf{v})$ in Eq.~\eqref{GNLaw}, yet are known to fall outside the standard Lagrangian formulation \cite{Neimark1972}. 
For example, consider a rigid disk rolling without slipping on a plane. The corresponding non-holonomic constraints,
\begin{eqnarray}
	\dot{x}=R\omega\cos\theta,
	\qquad
	\dot{y}=R\omega\sin\theta,
\end{eqnarray}
relate translational and rotational velocities and are therefore explicitly velocity dependent. The associated static-friction reaction force enforcing the rolling constraint $\mathbf{F}_{\mathrm{constraint}}$, performs no work and satisfies
\begin{eqnarray}
	\mathbf{F}_{\mathrm{constraint}}\cdot\mathbf{v}=0\,.
\end{eqnarray}
Consequently, it can always be represented in the transverse form
\begin{eqnarray}
	\mathbf{F}_{\mathrm{constraint}}=\mathbf{v}\times\mathbf{K}(\mathbf{x},\mathbf{v})\,,
\end{eqnarray}
with
\begin{eqnarray}
	\mathbf{K}(\mathbf{x},\mathbf{v})
	=
	\frac{\mathbf{F}_{\mathrm{constraint}}\times\mathbf{v}}{v^2}\,.
\end{eqnarray}
Since the rolling constraints depend explicitly on the velocities, the resulting function $\mathbf{K}$ is generally velocity dependent and cannot, in general, be reduced to the restricted form $\mathbf{K}=\nabla\times\mathbf{A}(\mathbf{x})$ obtained in Eq.~\eqref{CrossProdM2}. Therefore, such systems belong naturally to the broader PCE class while lying outside the standard PLA structure considered here.

Additional candidate classes include systems described within Nambu mechanics \cite{Nambu1973,Takhtajan1994}, where the dynamics is generated by multiple Hamiltonians and exhibits force structures that extend beyond the conventional Lagrangian framework.

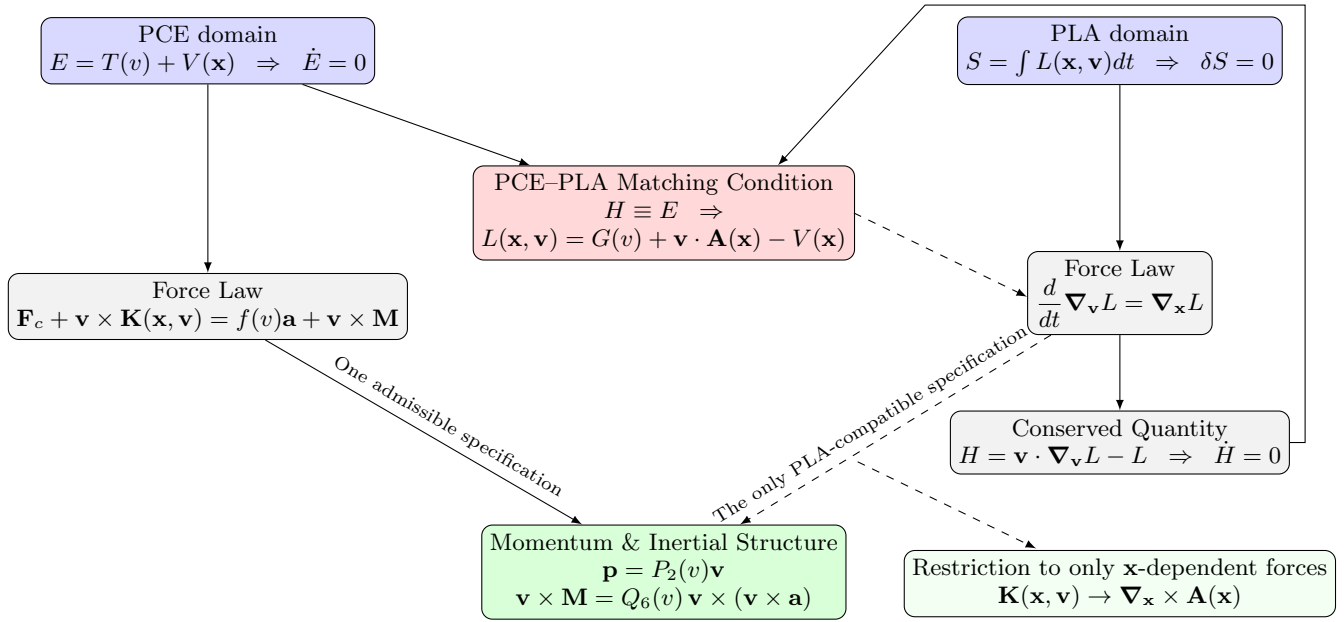
\begin{figure}[t]
\centering
\begin{tikzpicture}[
    >=latex,
    node distance=10mm and 35mm,
    boxpce/.style={rectangle,rounded corners,draw,fill=blue!15,align=center,inner sep=3pt,font=\small},
    boxres/.style={rectangle,rounded corners,draw,fill=gray!10,align=center,inner sep=3pt,font=\small},
    boxmatch/.style={rectangle,rounded corners,draw,fill=green!15,align=center,inner sep=3pt,font=\small},
    every path/.style={->}
]

\node[boxpce] (pce0) at (0,0)
  {PCE domain\\$\ds E=T(v)+V(\mathbf{x}) \;\;\Rightarrow\;\; \dot{E}=0$};

\node[boxres,below=25mm of pce0] (pce1)
  {Force Law\\ $\mathbf{F}_c + \mathbf{v}\times\mathbf{K}(\mathbf{x},\mathbf{v})
  = f(v)\mathbf{a} + \mathbf{v}\times \mathbf{M}$};

\node[boxpce] (pla0) at (12,0)
  {PLA domain\\ $S=\int L(\mathbf{x},\mathbf{v})\md t \;\;\Rightarrow\;\; \delta S = 0$};

\node[boxres,below=22mm of pla0] (pla1)
  {Force Law\\ $\dfrac{d}{dt}\boldsymbol{\nabla}_\mathbf{v} L
  = \boldsymbol{\nabla}_\mathbf{x} L$};

\node[boxres,below=of pla1] (pla2)
  {Conserved Quantity\\
   $H = \mathbf{v}\cdot\boldsymbol{\nabla}_\mathbf{v} L - L \;\;\Rightarrow\;\; \dot{H}=0$};

\node[boxres,below=of pla2, fill=green!5] (pla3)
  {Restriction to only $\mathbf{x}$-dependent forces\\
   $\mathbf{K}(\mathbf{x},\mathbf{v})\to\boldsymbol{\nabla}_\mathbf{x}\times\mathbf{A}(\mathbf{x})$};

\coordinate (leftmid)  at ($(pce0)!0.5!(pce1)$);
\coordinate (rightmid) at ($(pla0)!0.5!(pla1)$);

\node[boxmatch, fill=red!15, yshift=-5mm] (match1) at ($(leftmid)!0.5!(rightmid)$)
  {PCE--PLA Matching Condition \\ $H \equiv E \;\;\Rightarrow$\\
   $L(\mathbf{x},\mathbf{v}) = G(v) + \mathbf{v}\cdot\mathbf{A}(\mathbf{x}) - V(\mathbf{x})$};

\node[boxmatch, yshift=-25mm, below=of match1] (pmcor)
  {Momentum \& Inertial Structure\\
   $\mathbf{p}=P_2(v)\mathbf{v}$ \\
   $\mathbf{v}\times\mathbf{M}=Q_6(v)\,\mathbf{v}\times(\mathbf{v}\times\mathbf{a})$};

\draw (pce0) -- (pce1);
\draw (pce1) -- (pmcor)
    node[pos=0.55,sloped,above] {\scriptsize One admissible specification};

\draw (pla0) -- (pla1);
\draw (pla1) -- (pla2);

\draw (pce0) -- (match1);
\draw (pla2.east) 
      -- ++(2mm,0)                 
      -- ++(0,58mm)                
      -- ++(-50mm,0)                
       -- ($(match1.north)+(15mm,0)$);           

\draw[->,dashed] (8.55,-5.45) -- (10.9,-6.6);

\draw[dashed] ($(match1)+(2.5,0.0)$) --
      ($(pla1)+(-12.2mm,-0.5mm)$);
\draw[dashed] (pla1) -- ($(pmcor) + (10mm,6.3mm)$)
      node[pos=0.55,sloped,above] {\scriptsize The only PLA-compatible specification};

\end{tikzpicture}
\caption{Roadmap of the structural relations between PCE and PLA.
The functions $G$, $P_2$, and $Q_6$ are determined by the scalar response
$f(v)$, set by the minimal invariance condition in Eq.~\eqref{Invar00}.}
\label{Fig2}
\end{figure}

Having clarified the connection between PCE and PLA, as a closing remark, we conclude with a brief comparison between the momentum definitions stemming from each approach and their physical relevance.  The mechanical (or kinetic) momentum $\mathbf{p}(\mathbf{v})$, defined through the energy-based formulation via $\mathbf{F}_{\mathrm{tot}} = \md \mathbf{p}/\md t$, represents the intrinsic inertial response of the particle and depends only on its state of motion. In contrast, the canonical momentum $\mathbf{p}^\mtn{\mathrm{can}}(\mathbf{x},\mathbf{v}) = \boldsymbol{\nabla}_{\mathbf{v}} L(\mathbf{x},\mathbf{v}) = \mathbf{p}(\mathbf{v}) + \mathbf{A}(\mathbf{x})$, includes an additional contribution arising from the coupling to the external vector field $\mathbf{A}(\mathbf{x})$.
This distinction is familiar from electromagnetism, where $\mathbf{A}(\mathbf{x})$ corresponds to the magnetic vector potential. In this case, the mechanical (gauge-invariant) momentum is associated with the particle’s motion, while the canonical momentum incorporates the effect of the field coupling. The Lorentz force law depends only on the mechanical momentum, whereas the canonical momentum generates spatial translations and enters the Hamiltonian formulation.
In summary, mechanical momentum characterizes the particle’s inertial content and governs its response to forces, whereas canonical momentum encodes the combined particle–field system and serves as the generator of translations within the variational framework. The two coincide only in the absence of velocity–position couplings $(\mathbf{A}=\mathbf{0})$.

Fig. \ref{Fig2} provides a graphical roadmap of the correspondence between the PCE and PLA formulations.

\section{Conclusions}\label{Conl}
In this work we have developed a formulation of particle mechanics in which the functional relation between force and kinetic energy is derived directly from local mechanical energy conservation rather than postulated through Newton’s second law or a variational principle. By imposing the instantaneous condition $\dot{E}=0$ as a pointwise constraint along a particle trajectory, we obtained a generalized force law without assuming a specific kinetic-energy function, momentum–velocity relation, or predefined equation of motion.

A central outcome of this approach is, on the one hand, the accommodation of both conservative and non-conservative (yet non-dissipative) forces, and on the other hand, the natural decomposition of the inertial response into two distinct components: a collinear component with the acceleration and responsible for changes in kinetic energy, and a transverse component, orthogonal to the velocity, which preserves energy while modifying the direction of motion. This structure emerges solely from local energy balance and persists independently of the particular realization of the dynamics. Imposing rotational equivariance further constrains the admissible form of the force law, ensuring compatibility with spatial isotropy.

We showed that in one-dimensional motion, the equivalence of inertial reference frames implies invariance of the total force under inertial boosts. In one dimension, this invariance, together with local energy conservation, uniquely fixes the functional forms of kinetic energy and momentum. Galilean invariance selects the standard Newtonian expressions, while Lorentz invariance yields the relativistic ones. Newtonian and relativistic mechanics thus emerge as symmetry-selected realizations of a common underlying force–energy structure, rather than as fundamentally distinct dynamical theories.

By comparing the energy-based derivation with the principle of least action, we identified the precise structural conditions under which variational formulations reproduce the same dynamics as those obtained from local energy conservation. This analysis clarifies the domain of equivalence between the two approaches and highlights the broader class of energy-conserving dynamics that lie beyond the standard variational framework, subject though to its principal limitations, including the restriction to single-particle systems, the separability assumption $E=T(v)+V(\mathbf{x})$, and the non-covariance of the relativistic force law.

Taken together, these results show that local energy conservation provides a powerful and conceptually transparent foundation for particle mechanics. Within this perspective, force laws, kinetic energy, and momentum are not independent postulates but interdependent structures constrained jointly by energetic balance and inertial-frame symmetry.

Future work may explore the role of transverse, energy-preserving responses in non-inertial frames or extended systems, identify concrete physical realizations of the acceleration-coupled transverse response and the experimental conditions under which it may become observable, and generalize the formalism to higher-dimensional settings with appropriate replacements for the cross product. Further directions include extending the framework to many-particle systems, where collective and interaction effects may qualitatively modify the structure of the energy balance, as well as to situations in which the mechanical energy depends explicitly on time dependence too.
A particularly important direction is to identify explicit systems whose dynamics satisfy the Principle of Energy Conservation while falling outside any standard variational formulation, thereby providing a concrete theoretical or empirical criterion for distinguishing between these two formulations of dynamics. Finally, it would be of considerable interest to investigate how the present framework interfaces with field-theoretic formulations and whether analogous energy-based structures persist in quantum or semiclassical contexts.

\begin{acknowledgments}
The author would like to thank anonymous Referee 2 and Referee Jacob Nagler for constructive comments and suggestions that helped improve the manuscript.
\end{acknowledgments}

\appendix

\section{Proof of Eq. \eqref{MomExp}}\label{AppendixA0}

In this appendix we present the necessary intermediate steps leading to the expressions in Eq.~\eqref{MomExp}. Imposing $SO(3)$-equivariance on the vectors $\mathbf{p}$ and $\mathbf{M}$ in Eq.~\eqref{MomentumDecomp} directly yields Eqs.~\eqref{Conditions}–\eqref{SecIIeq10}. Substituting Eq.~\eqref{SecIIeq10} into Eq.~\eqref{Conditions}, we obtain
\begin{subequations}\label{AppA0eq1}
\begin{eqnarray}
\label{AppA0eq1_a}
\dot{P}_1 &=& Q_4v^2 + Q_5(\mathbf{v}\cdot\mathbf{a})\,,\\
\label{AppA0eq1_b}
P_1 + \dot{P}_2 &=& -Q_4(\mathbf{x}\cdot\mathbf{v}) + Q_6(\mathbf{v}\cdot\mathbf{a})\,,\\
\label{AppA0eq1_c}
P_2 &=& f(v) - Q_5(\mathbf{x}\cdot\mathbf{v})-Q_6v^2\,.
\end{eqnarray}
\end{subequations}
Taking the time derivative of $P_2$ from Eq.~\eqref{AppA0eq1_c}, we find
\begin{eqnarray}
\dot{P}_2 &=& 
\dot{f}(v) 
- \dot{Q}_5(\mathbf{x}\cdot\mathbf{v})
- Q_5 \left[v^2+ (\mathbf{x}\cdot\mathbf{a}) \right]
- \dot{Q}_6v^2 - 2Q_6 (\mathbf{v}\cdot\mathbf{a})\,,
\end{eqnarray}
which, upon substitution into Eq.~\eqref{AppA0eq1_b}, allows us to determine $P_1$ as
\begin{eqnarray}\label{SecIIeq3}
P_1
=
-\dot{f}(v)+(\dot{Q}_5-Q_4)(\mathbf{x}\cdot\mathbf{v})+Q_5(\mathbf{x}\cdot\mathbf{a}) + 3Q_6(\mathbf{v}\cdot\mathbf{a})+(Q_5+\dot{Q}_6)v^2\,.
\end{eqnarray}
Consistency requires that the time derivative of Eq.~\eqref{SecIIeq3} reproduces Eq.~\eqref{AppA0eq1_a}. Enforcing this condition gives
\begin{eqnarray}
\nonumber
\ddot{f}(v)
&=&
\frac{\md}{\md t}
[(\dot{Q}_5-Q_4)(\mathbf{x}\cdot\mathbf{v})+Q_5(\mathbf{x}\cdot\mathbf{a}) + 3Q_6(\mathbf{v}\cdot\mathbf{a})+(Q_5+\dot{Q}_6)v^2]
- Q_4v^2 
- Q_5(\mathbf{v}\cdot\mathbf{a})\\
\nonumber
&=&
\frac{\md}{\md t}
\left[(\dot{Q}_5-Q_4)(\mathbf{x}\cdot\mathbf{v})+Q_5(\mathbf{x}\cdot\mathbf{a})\right]
+
\frac{\md}{\md t}
\left[3Q_6(\mathbf{v}\cdot\mathbf{a})+(Q_5+\dot{Q}_6)v^2\right]
- Q_4v^2 
- Q_5(\mathbf{v}\cdot\mathbf{a})\\
\nonumber
&=&
\mathbf{x}\cdot\frac{\md}{\md t}
\left[(\dot{Q}_5-Q_4)\mathbf{v}+Q_5\mathbf{a}\right]
+
(\dot{Q}_5-Q_4)v^2
+
\frac{\md}{\md t}
\left[3Q_6(\mathbf{v}\cdot\mathbf{a})+(Q_5+\dot{Q}_6)v^2\right]
- Q_4v^2 \\
\nonumber
&=&
\mathbf{x}\cdot\frac{\md}{\md t}
\left[(\dot{Q}_5-Q_4)\mathbf{v}+Q_5\mathbf{a}\right]
+
2 Q_5 (\mathbf{v}\cdot\mathbf{a})
+
2\left(\dot{Q}_5-Q_4\right)v^2
+
\frac{\md}{\md t}
\left[3Q_6(\mathbf{v}\cdot\mathbf{a})+\dot{Q}_6v^2\right]\\
&=&
\mathbf{x}\cdot 
\frac{\md}{\md t}
\left[ 
\left(\dot{Q}_5-Q_4\right)\mathbf{v}
+ 
Q_5\mathbf{a}
\right]
+
2Q_5(\mathbf{v}\cdot\mathbf{a})  
+ 
2\left(\dot{Q}_5-Q_4\right)v^2 
+
\frac{\md}{\md t} \left[Q_6 (\mathbf{v}\cdot\mathbf{a}) + \frac{\md}{\md t}(v^2 Q_6)\right]
\end{eqnarray}
This expression coincides with Eq.~\eqref{Constaint1}, which couples the kinetic energy functional $f$ to the coefficients appearing in the response vector $\mathbf{M}$.

To eliminate any explicit dependence on position, we set $Q_4 = Q_5 = 0$, reducing the coupling relation to
\begin{eqnarray}
\ddot{f}(v)
&=&
\frac{\md}{\md t} \left[Q_6 (\mathbf{v}\cdot\mathbf{a}) + \frac{\md}{\md t}(v^2 Q_6)\right]
=
\frac{\md}{\md t} \left[Q_6 v\frac{\md v}{\md t} + \frac{\md}{\md t}(v^2Q_6)\right]\,.
\end{eqnarray}
Integrating w.r.t. time yields
\begin{eqnarray}
f'(v)\frac{\md v}{\md t}
=
\dot{f}(v)
=
Q_6 v\frac{\md v}{\md t} + \frac{\md}{\md t}(v^2Q_6)+\alpha_1\,,
\end{eqnarray}
or equivalently
\begin{eqnarray}
\left[f'(v)-Q_6 v-\frac{\md}{\md v}(v^2Q_6)\right]\frac{\md v}{\md t}
=
\alpha_1\,,
\end{eqnarray}
where we have used the chain rule $\dot{H}(v)=H'(v),\frac{\md v}{\md t}$ for a differentiable function $H(v)$ and sufficiently smooth $v(t)$. Setting the integration constant $\alpha_1=0$, the relation holds for arbitrary $\frac{\md v}{\md t}$ only if
\begin{eqnarray}
f'(v)-Q_6 v-\frac{\md}{\md v}(v^2Q_6)=0\,.
\end{eqnarray}
Multiplying by $v \neq 0$ gives
\begin{eqnarray}
vf'(v)
=
v^2Q_6+v\frac{\md}{\md v}(v^2Q_6)
=
\frac{\md}{\md v}(v^3 Q_6)\,,
\end{eqnarray}
which can be integrated to obtain
\begin{eqnarray}
Q_6(v)
&=&
\frac{1}{v^3}\left(\int v \frac{\partial f(v)}{\partial v}\md v - \alpha\right)\,,
\end{eqnarray}
with integration constant $\alpha$, as stated in Eq.~\eqref{ImpDiffEq}. Substituting this result into the relevant expressions immediately yields Eq.~\eqref{MomExp}.

\section{Proof of Eq. \eqref{Invar00}}\label{AppendixA}

In this appendix we prove that, for one–dimensional motion, the transformation law for the total force between two IRFs related by a boost of velocity $u$ must be of the identity form.  

For each fixed boost velocity $u\in\mathbb{R}$ we denote by $\Phi_u:\mathbb{R}\to\mathbb{R}$ the map that assigns to a force $F$ measured in frame $\Sigma$ the corresponding force $F'$ measured in frame $\Sigma'$:
\[
F'=\Phi_u(F)\,.
\]
Our goal is to show that necessarily
\[
\Phi_u(F)=F\,,\qquad \text{for all } u\,.
\]

Before proceeding, we emphasize that the argument applies to 1D motion and to the total force acting on a point particle, under the standard assumptions that i. IRFs are related by smooth boosts, and ii. forces satisfy the usual superposition principle within each IRF. Situations involving velocity-dependent interaction laws with intrinsic direction-dependent structure (e.g., genuinely higher-dimensional effects) fall outside the scope of this derivation.

To this end we list the structural properties that $\Phi_u$ must satisfy as a consequence of the Relativity Principle together with the standard kinematics of IRFs:
\begin{enumerate}[label=(\roman*),ref=(\roman*)]
\item \textit{Additivity (superposition):} If two interactions act independently on the same particle, their total force is the sum $F_1+F_2$ in any given IRF. Since the decomposition into independent interactions is a physical statement (not frame-dependent), transforming to another IRF must preserve this linear superposition:
        \[
            \Phi_u(F_1+F_2)=\Phi_u(F_1)+\Phi_u(F_2)
            \qquad \forall F_1,F_2\in\mathbb{R}.
        \]
        \label{prop:additivity}
This assumption encodes the empirical fact that forces combine linearly and that this rule is frame-independent.
\item \textit{Order preservation:}  In one dimension, the sign of $F$ specifies the direction of acceleration. A change of inertial frame should not reverse this direction discontinuously for arbitrarily small boosts. Accordingly, for each fixed $u$, the map $\Phi_u$ is assumed to be monotone in $F$. Physically, this excludes pathological transformations in which a right-pointing force in one frame systematically appears left-pointing in another without a corresponding reversal of spatial orientation. \label{prop:monotonicity}
\item \textit{Inverse–boost property:} Boosting from $\Sigma$ to $\Sigma'$ with velocity $u$ and then back with velocity $-u$ must return the original physical description:
        \[
            \Phi_{-u}\!\left(\Phi_u(F)\right)=F \qquad \forall F\in\mathbb{R}.
        \]
        Thus $\Phi_u$ is invertible and its inverse is $\Phi_{-u}$. \label{prop:inverse}
\item \textit{Identity at zero relative velocity:} If both IRFs $\Sigma $ and $\Sigma'$ are stationary, $u=0$, then
        \[
            \Phi_0(F)=F\,,
        \]
        since a vanishing relative velocity corresponds to no transformation. \label{prop:identity}

    \item \textit{Continuity (or smoothness) in the boost parameter:} the measured force in $\Sigma'$ should depend continuously on the relative velocity $u$. Thus, for each fixed $F$, the map $u\mapsto \Phi_u(F)$ is assumed to be smooth. This excludes discontinuous or singular frame transformations. \label{prop:continuity}
\end{enumerate}

Property~\ref{prop:additivity} is the classical Cauchy functional equation \cite{Hamel1905,Aczel1966}. A standard result from real analysis states that if a function
$\Phi:\mathbb{R}\to\mathbb{R}$ satisfies
\[
\Phi(F_1+F_2)=\Phi(F_1)+\Phi(F_2)
\]
and is either continuous or monotone, then it must be linear:
\[
\Phi(F)=cF \qquad \text{for all }F\in\mathbb{R},
\]
where $c=\Phi(1)$.
Because Property~\ref{prop:monotonicity} ensures monotonicity in $F$, respectively, we conclude that for each fixed $u$
\[
\Phi_u(F)=k(u)\,F
\]
for some real function $k(u)$.

Since $k(u)$ is continuous in $u$ (Property \ref{prop:continuity}) and satisfies $k(0)=1>0$ (Property \ref{prop:identity}), any change of sign would, by the intermediate value theorem, require the existence of some $u_c$ such that $k(u_c)=0$. However, this is incompatible with the inverse–boost condition $k(u)k(-u)=1$ (Property \ref{prop:inverse}), which excludes zeros for all $u\in\mathbb{R}$. It follows that $k(u)>0$ for all $u\in\mathbb{R}$, and then the product constraint implies $k(u)=1$ for all $u\in\mathbb{R}$.

Accordingly, for any relative speed $u$ in 1D motion, the total force $F$ acting on a particle is not only form–invariant (covariant) but also value–invariant in any other IRF:
\[
F'=\Phi_u(F)=F\,,
\]
or equivalently
\[
f(v)a=f(v')a'\,.
\]

The result hinges crucially on the combination of linear superposition, monotonicity, and smooth dependence on the boost parameter. Relaxing any of these assumptions (for instance, allowing non-additive interactions or discontinuous transformations) would open the possibility of more general transformation laws.

\section{Proof of Eq. (\ref{eq41})}\label{AppendixB}

This appendix provides the technical derivation of Eq. (\ref{eq41}) used in Section \ref{SecIV}. Throughout this derivation, it is assumed that $v\neq0$, so that divisions by $v$ are well defined.

We begin by expanding the term $\boldsymbol{\nabla}_{\mathbf{v}}G(v)$, namely
\begin{eqnarray}
\boldsymbol{\nabla}_{\mathbf{v}}G(v)
=
\frac{\partial G(v)}{\partial v} \boldsymbol{\nabla}_{\mathbf{v}}v
=
\frac{1}{v}\frac{\partial G(v)}{\partial v} \mathbf{v}\,.
\end{eqnarray}
Then, its time derivative yields
\begin{eqnarray}
\frac{\md}{\md t}\boldsymbol{\nabla}_{\mathbf{v}}G(v)
\nonumber
&=&
\left(\frac{\md}{\md t}\left[\frac{1}{v}\frac{\partial G(v)}{\partial v}\right]\right) \mathbf{v}
+
\frac{1}{v}\frac{\partial G(v)}{\partial v} \mathbf{a}\\
\nonumber
&=&
\left( -\frac{1}{v^2}\frac{\md v}{\md t} \frac{\partial G(v)}{\partial v} + \frac{1}{v}\frac{\md}{\md t}\frac{\partial G(v)}{\partial v} \right) \mathbf{v}
+
\frac{1}{v}\frac{\partial G(v)}{\partial v} \mathbf{a}\\
&=&
\left( 
-\frac{1}{v^2}\frac{\md v}{\md t} \frac{\partial G(v)}{\partial v} 
+ 
\frac{1}{v}\frac{\md v}{\md t}\frac{\partial^2 G(v)}{\partial v^2} 
\right) \mathbf{v}
+
\frac{1}{v}\frac{\partial G(v)}{\partial v} \mathbf{a}\\
\nonumber
&=&
\left(
-\frac{1}{v^3}\left(v\frac{\md v}{\md t}\right) \frac{\partial G(v)}{\partial v} 
+ 
\frac{1}{v^2}\left(v\frac{\md v}{\md t}\right)\frac{\partial^2 G(v)}{\partial v^2} 
\right) \mathbf{v}
+
\frac{1}{v}\frac{\partial G(v)}{\partial v} \mathbf{a}\\
\nonumber
&=&
\left(
-\frac{1}{v} \frac{\partial G(v)}{\partial v} 
+ 
\frac{\partial^2 G(v)}{\partial v^2} 
\right) \frac{(\mathbf{v}\cdot\mathbf{a})}{v^2}\mathbf{v}
+
\frac{1}{v}\frac{\partial G(v)}{\partial v} \mathbf{a}\,.
\end{eqnarray}
Substituting this result into Eq. (\ref{CrossProdM1}), and taking into account the definition of $f$ in Eq. (\ref{fdef}), we have
\begin{eqnarray}\label{AppB3}
\mathbf{v}\times\mathbf{M}
&=&
\left(
\frac{\partial^2 G(v)}{\partial v^2} -\frac{1}{v} \frac{\partial G(v)}{\partial v} 
\right) \frac{(\mathbf{v}\cdot\mathbf{a})}{v^2}\mathbf{v}
+
\frac{1}{v}\left( \frac{\partial G(v)}{\partial v} - \frac{\partial T(v)}{\partial v} \right)\mathbf{a}
\end{eqnarray}
Differentiating Eq. (\ref{ExplLang}) we obtain the relation $T=v G'- G$. Substituting this expression for 
$T$ into Eq.~\eqref{AppB3} yields
\begin{eqnarray}
\mathbf{v}\times\mathbf{M}
\nonumber
&=&
\left(
\frac{\partial^2 G(v)}{\partial v^2} -\frac{1}{v} \frac{\partial G(v)}{\partial v} 
\right) \frac{(\mathbf{v}\cdot\mathbf{a})}{v^2}\mathbf{v}
-
\left(\frac{\partial^2 G(v)}{\partial v^2} -\frac{1}{v}\frac{\partial G(v)}{\partial v}\right)\mathbf{a}\\
&=&
\frac{1}{v^2}\left(
\frac{\partial^2 G(v)}{\partial v^2} - \frac{1}{v} \frac{\partial G(v)}{\partial v}  
\right) 
\Big( (\mathbf{v}\cdot\mathbf{a})\mathbf{v}-v^2\mathbf{a}\Big)\\
\nonumber
&=&
\frac{1}{v^2}\left(
\frac{\partial^2 G(v)}{\partial v^2} - \frac{1}{v} \frac{\partial G(v)}{\partial v}  
\right) 
\mathbf{v}\times(\mathbf{v}\times\mathbf{a})\,.
\end{eqnarray}
Re-expressing $G$ in terms of $f$, we see that
\begin{eqnarray}
\frac{1}{v^2}\left(
\frac{\partial^2 G(v)}{\partial v^2} - \frac{1}{v} \frac{\partial G(v)}{\partial v}  
\right) 
=
Q_6(v)\,,
\end{eqnarray}
where $Q_6$ is given in Eq. \eqref{ImpDiffEq}. This completes the derivation of Eq. \eqref{eq41}, as used in Section \ref{SecIV}.


\end{document}